\documentclass[aps,prb,superscriptaddress,longbibliography,twocolumn]{revtex4-1}

\usepackage{setspace}
\usepackage{lmodern}

\usepackage[latin9]{inputenc}
\usepackage{amsmath}
\usepackage{amssymb}
\usepackage{graphicx}
\usepackage{epstopdf}
\usepackage{color}
\usepackage{enumerate}
\usepackage{array}
\usepackage{tabularx}
\usepackage{mathtools} 
\usepackage{multirow}
\usepackage{float}
\begin{document}

\title{A tunable Fabry-P\'{e}rot quantum Hall interferometer in graphene}

\author{Corentin D\'{e}prez}
\author{Louis Veyrat}
\author{Hadrien Vignaud}
\author{Goutham Nayak}
\affiliation{Univ. Grenoble Alpes, CNRS, Grenoble INP, Institut N\'{e}el, 38000 Grenoble, France}
\author{Kenji Watanabe}
\affiliation{National Institute for Materials Science, 1-1 Namiki, Tsukuba 306-0044, Japan}
\author{Takashi Taniguchi}
\affiliation{International Center for Materials Nanoarchitectonics}
\author{Fr\'{e}d\'{e}ric Gay}
\author{Hermann Sellier}
\author{Benjamin Sac\'{e}p\'{e}}
\email{Corresponding author : benjamin.sacepe@neel.cnrs.fr}
\affiliation{Univ. Grenoble Alpes, CNRS, Grenoble INP, Institut N\'{e}el, 38000 Grenoble, France}

\begin{abstract}
\textbf{Electron interferometry with quantum Hall edge channels holds promise for probing and harnessing exotic exchange statistics of non-Abelian anyons. 
In semiconductor heterostructures, however, quantum Hall interferometry has proven challenging and often obscured by charging effects. Here we show that  high-mobility monolayer graphene equipped with a series of gate-tunable quantum point contacts that act as electron beam-splitters provides a model system to perform Fabry-P\'{e}rot quantum Hall interferometry. We observe high-visibility Aharonov-Bohm interference free of charging effects and widely tunable through electrostatic gating or magnetic field, in remarkable agreement with theory. A coherence length of $\mathbf{10 \,\mu m}$ at a temperature of $0.02$ K allows us to further achieve coherently-coupled double Fabry-P\'{e}rot interferometry. Our results open a new avenue for quantum Hall interferometry and the exploitation of topological excitations for quantum computation.}
\end{abstract}

\maketitle

The wave-like behavior of electrons is exemplified in metals and semiconductors by a variety of mesoscopic phenomena stemming from quantum interference effects.
Universal conductance fluctuations in coherent conductors or quantum (weak) localization due to random scattering are some vivid examples.
In two dimensional electron gases, quantum interferences can be precisely tailored and harnessed using the chiral edge channels of the paradigmatic quantum Hall (QH) effect as one-dimensional, coherent electron beams.  
Electrostatic manipulation and partitioning of QH edge channel trajectories via local gates and quantum point contacts (QPCs) --that is, tunable beam-splitters for electrons-- makes it possible to construct on-chip electronic analogues of Fabry-P\'{e}rot~\cite{Wees89b} (FP) or Mach-Zehnder~\cite{Ji03} optical interferometers, usable for quantum information processing~\cite{bauerle18}.

Central to QH interferometry are the braiding of anyonic excitations~\cite{Chamon1997} in the fractional QH regime and the prospect of exploiting the non-Abelian properties~\cite{Fradkin98,DasSarma05,Bonderson06a,Chung06,Stern06,Feldman06,Stern10} of some fractional QH states~\cite{Banerjee18} for topological quantum computation~\cite{Nayak08}. Winding fractional edge excitations around localized bulk ones is the elementary process sensitive to the anyonic statistical exchange phase~\cite{Chamon1997} --a quantity evidenced via shot-noise measurements~\cite{Bartolomei20}.  Such a braiding process of anyonic excitations has been demonstrated very recently in Fabry-P\'{e}rot interferometers~\cite{Nakamura20} in which an edge channel, partially reflected between two QPCs, encloses localized excitations into an interfering loop  and picks up the anyonic statistical phase. Moreover, advanced devices based on coherently-coupled double Fabry-P\'{e}rot interferometers have been envisioned as a possible route for achieving topologically protected qubits addressable via braiding operations~\cite{DasSarma05}.

The body of work on QH Fabry-P\'{e}rot interferometers performed in GaAs heterostructures~\cite{Camino07,Zhang2009,McClure2009,Ofek2010,McClure2012,Willett13,Choi2015,Sivan2016,Nakamura2019} has, however, faced various difficulties.
Observing quantum interference tunable by gate electrodes and magnetic field through the Aharonov-Bohm effect has proven often disguised by charging effects in the Fabry-P\'{e}rot cavity~\cite{Camino07,Zhang2009,McClure2009,Ofek2010,McClure2012,Sivan2016,Roosli20}. 
The Coulomb blockade~\cite{Rosenow07,Halperin2011,NgoDinh2012} that is due to the small interferometer sizes has long hindered progress and demands device architectures that implement various types of screening electrodes to mitigate charging effects~\cite{Choi2015,Nakamura2019,Nakamura20}. 
Furthermore, the delicate edge channel reconstructions in GaAs involving neutral modes or counter-propagating charge modes~\cite{Lafont19} can complicate interferometry~\cite{Bhattacharyya19}. 
All these highlight the need for exploring QH interferometry in other two-dimensional electron gases.

Here, we demonstrate that graphene monolayer is a promising alternative platform for QH Fabry-P\'{e}rot interferometry, free of charging effects, thereby opening a new pathway for anyon interferometry~\cite{Nakamura2019,Nakamura20}. Graphene comes forth as an ideal two-dimensional electron gas with all necessary features: high-mobility~\cite{Wang13}, fractional quantum Hall effect~\cite{Du09,Bolotin09} with large energy gaps~\cite{Dean11}, including possible non-Abelian states~\cite{Kim2018,Zibrov2018}, helical edge channels at charge neutrality~\cite{Young14,Veyrat20}, gate-tunability, and versatility of van der Waals heterostructures~\cite{Yankowitz19}. Similarly for the graphene bilayer~\cite{Maher13,Ki14,Li17}.

Yet, gate tunable QPCs --the key component to construct elaborated QH interferometers (see Fig.~1)-- has proven challenging to realize due to the gapless band structure of graphene. Depleting the electron gas through electrostatic gating accumulates holes and yields conducting pn-junctions~\cite{Huard07} that short-circuit QPC constrictions~\cite{Veyrat19}. Other alternatives involve partitioning at a pn junctions~\cite{Wei17,Makk18}, or etched constrictions~\cite{Zhang19} that yield devices subject to charging effects.
We remedy this with the use of high-mobility graphene in the QH regime for which an insulating, broken-symmetry state develops and opens an energy gap at the charge neutrality point separating the conduction band from the valance band~\cite{Young14}. This gap is essential for the functioning of split gates as full-fledged QPC with tunable edge channel transmission~\cite{Zimmermann2017}: It suppresses some possible charge transfers at the pn junction surrounding the split-gates (see drawing in Fig.~2d).

We thus devised our device on the basis of a high-mobility graphene van der Waals heterostructure made with an hexagonal boron nitride (hBN)-encapsulated graphene monolayer~\cite{Wang13} resting atop a graphite flake that acts as a back-gate electrode. As depicted in Fig.~1a, three pairs of split-gates of $20$~nm gap, on top of the heterostructure, serve as QPCs and define a double FP cavity. Plunger gates located on the edges of each cavity are used to move the trajectory of the interfering QH edge channel and modulate the effective interferometer areas. The device we focus on in this report is shown in Fig.~1b with QPCs highlighted in red and labeled QPC$_1$, QPC$_2$ and QPC$_3$, plunger gates in orange and ohmic contacts in yellow (see Supp. Section III for the design characteristics of the QPC geometry). 
This triple QPC configuration enables us to operate three FP interferometers that we denote small (QPC$_2$-QPC$_3$), medium (QPC$_1$-QPC$_2$) and large (QPC$_1$-QPC$_3$) interferometers, whose geometrical areas are 3.1, 10.7 and 14.7 $\mu\text{m}^2$, respectively (see Supp. Table~S2).

To operate the QH-FP interferometers, the graphene electron gas is subjected to a perpendicular magnetic field, $B$, to reach the quantum Hall regime. Prior to electron interferometry, each QPC transmission is individually characterized by mapping out the diagonal resistance $R_{\text{D}}$ measured across contacts $V_{\text{D}}^+$ and $V_{\text{D}}^-$ as a function of back-gate and split-gate voltages (see Supp. Fig.~S5). 
We choose to work at bulk filling fractions $\nu_{\text{b}} = n_{\text{b}}\phi_0/B$  ($n_{\text{b}}$ is the bulk charge carrier density, $\phi_0=h/e$ is the flux quantum with $h$ the Planck constant and $e$ the electron charge) such that two electron-type edge channels of the zeroth Landau level propagate on the graphene edges, as indicated by a Hall conductance of $2e^2/h$ (see Supp. Fig.~S7). A proper tuning of the split-gate voltages enables us to partially transmit either the inner or outer edge channel at the QPCs. Other configurations with only one electron-type edge channel present in the bulk have also been studied (see Supp. Section XIII).\\

\textbf{Widely tunable quantum interference} 

The quantum interferences in the small electronic Fabry-P\'{e}rot interferometer readily show up in Fig.~2a-c, which display the oscillatory behavior of the diagonal resistance $R_{\text{D}}$ as a function of the plunger-gate voltage $V_{\text{pg2}}$. For this measurement performed at $B=14$~T, the QH-FP interferometer operates at $\nu_{\text{b}}= 1.5$. 
We tuned QPC$_2$ and QPC$_3$ to a transmission of the outer edge channel $T_2 = 0.60$ and  $T_3=0.47$ , respectively, leaving the inner one fully backscattered and localized in the FP cavity (see Supp. Fig.~S6 for the QPCs transmission curves). In this configuration, $R_D$ oscillates with about $50$\% visibility (see Fig.~2a and b) over nearly the whole voltage range spanned by the plunger gate (Fig.~2c), starting around $V_{\text{pg2}} =-0.3$~V down to $V_{\text{pg2}} =-4$~V. Further oscillations extending to positive $V_{\text{pg2}}$ values when electrons are accumulated beneath the plunger gate are shown in Supp. Fig.~S8. These results showing more than 280 resistance oscillations demonstrate the high stability and wide tunability of our interferometer. 

The oscillatory behavior of the diagonal resistance upon depleting the electron gas beneath the plunger gate is a direct consequence of the modulation $\delta A$ of the interferometer area  defined by the interfering edge channel. Changing the flux through the interferometer via  $\delta A$ yields a variation $\delta \varphi = 2\pi B\delta A / \phi_0$ of the Aharonov-Bohm phase picked up by the electrons. The flux-modulated succession of constructive and destructive electron wave-function interferences is thus reflected in the stark oscillations of the diagonal resistance.\\

\textbf{Plunger-gate electrostatics} 

Inspecting the oscillations more closely on a smaller span around two different plunger-gate voltages, $V_{\text{pg2}} =-3$~V and $-1$~V in Fig.~2a and b, respectively, we see that the period depends upon the plunger-gate voltage. This behavior can be tracked by computing the Fourier transform restricted to a small voltage window sliding over the entire $V_{\text{pg2}}$ range. The resulting Fourier amplitude shown in Fig.~2e as a function of $V_{\text{pg2}}$ and plunger-gate-voltage frequency $f_{\text{pg2}}$ displays a single peak that disperses to lower frequency upon decreasing  $V_{\text{pg2}}$ to more negative values. This is consistent with the larger periodicity observed in Fig.~2a with respect to that of Fig.~2b. Notice that a second harmonic indicated by the black arrow in the inset of Fig.~2e is also visible, mostly at large negative $V_{\text{pg2}}$.

The  $V_{\text{pg2}}$-dispersion of the oscillations frequency reflects the electrostatics of the plunger gate. 
Depleting the graphene electron gas and then accumulating hole states locally under the plunger gate repels the interfering edge channel towards the interior of the QH-FP interferometer as illustrated in the schematics in Fig.~2d, therefore reducing the effective area. The abrupt drop of the oscillation frequency $f_{\text{pg2}}$ at $V_{\text{pg2}} \simeq -0.3$~V corresponds to the expulsion out of the area beneath the plunger gate of the interfering edge channel that was initially propagating along the graphene edge. This expulsion occurs when the graphene beneath the plunger gate reaches charge neutrality (or filling fraction 1 for the inner edge channel interfering case, see Fig.~2d). Notice that near this regime, the apparent reduction of the oscillation amplitude is not physical but results from a too fast oscillation frequency with respect to the bandwidth of our measurement and the constant sweep rate of the plunger gate voltage. For more negative $V_{\text{pg2}}$, $f_{\text{pg2}}$ decreases more slowly due to the accumulation of holes and the ensuing displacement of the pn-junction further away from the plunger gate. 

The plunger gate tuning of the magnetic flux is demonstrated by the magnetic field dependence of the oscillations period.
The area variation $\Delta A = \phi_0/B$ that yields a change of one flux quantum in the interferometer relates to the plunger-gate voltage period $\Delta V_{\text{pg2}}$ by $\Delta A =\alpha \Delta V_{\text{pg2}}$, where $\alpha $ is the (non-linear) lever arm of the gate. As $f_{\text{pg2}}=1/ \Delta V_{\text{pg2}}$, the quantity $f_{\text{pg2}}/B = \alpha / \phi_0$ depends only on the electrostatic displacement of the pn interface that is encoded in $\alpha $. Fig.~2f displays $f_{\text{pg2}}/B$ as a function of  $\bar{V}_{\text{pg2}}$, gathered from Fig.~2e and from similar data obtained at $B=11$ and $8$~T, and also by making interfering the inner edge channel through a suitable tuning of the QPCs (see Supp. Fig.~S9). Here, $\bar{V}_{\text{pg2}}$ is the plunger-gate voltage shifted with respect to the voltage that expels the inner or outer edge channel. Despite a large variation of magnetic field, and hence of oscillations periods, all data collapse onto the same curve, confirming the flux periodicity. 

Furthermore, this data collapse draws the $\bar{V}_{\text{pg2}}$-evolution of the lever arm, which can be directly compared to numerical simulations of the electrostatic displacement of the pn junction in our device geometry (see Supp. Fig.~S10). The resulting computation of $\alpha /\phi_0$ shown by the black line in Fig.~2f consistently fits the data and therefore demonstrates the $\phi_0$-periodicity of the oscillations, bearing out the Aharonov-Bohm origin of the resistance oscillations.\\

\textbf{Aharonov-Bohm effect vs Coulomb blockade} 

A critical aspect of QH-FP interferometers lies in the possibility that the resistance oscillations result from charging effects in the FP cavity~\cite{Rosenow07,Halperin2011,NgoDinh2012}, instead of Aharonov-Bohm quantum interference. Thorough studies on this issue showed that these two competing phenomena can be straightforwardly differentiated by the magnetic-field dependence of the gate-induced oscillations~\cite{Zhang2009,Ofek2010,Halperin2011,NgoDinh2012,Sivan2016}: For the Aharonov-Bohm effect, resistance oscillations in the $B$--$V_{\text{pg}}$ plane shall draw diagonal lines of negative slope, indicating constant flux $\phi = B\delta A + A \delta B$ through the interferometer, whereas a zero or positive slope is expected for the Coulomb blockade effect. This led us to perform systematic measurements of the resistance oscillations as a function of plunger-gate voltage and magnetic-field variation $\delta B$. Figure~3 displays two typical resistance maps obtained with the small interferometer (Fig.~3a) and with the large one (Fig.~3b), both at $B=14$~T with the outer edge channel interfering. For both cases, resistance maxima (minima) draw lines of negative slope: Upon increasing magnetic field the lines go to more negative values of  $V_{\text{pg}}$, hence shrinking the area to maintain the Aharonov-Bohm phase (flux) constant.
This behavior that we constantly observed in all configurations, regardless of the interferometer size, magnetic field value, bulk filling factor being 1 or 2, or which edge channel is interfering (see Supp. Sections X and XI), definitely rules out the alternative Coulomb blockade scenario. 

The absence of Coulomb blockade even in the smaller interferometer, whose dimensions are similar to those in GaAs devices that exhibit Coulomb blockade~\cite{Zhang2009}, points to a specificity of the hBN-graphene heterostructure. For our devices, the main source of charging effect mitigation is the close proximity of the graphite back gate electrode. Following the theoretical approach of Ref.~\cite{Halperin2011}, we evaluated the various capacitances involved in our devices and calculated the parameter  $\xi = \frac{C_{\rm eb}}{C_{\rm b}+C_{\rm eb}}$ with  $C_{\rm b}$ the bulk-to-gate capacitance and $C_{\rm eb}$ the edge-to-bulk capacitance, which defines the Aharonov-Bohm ($\xi \ll1$) or Coulomb-dominated ($\xi \sim 1$) operating regime~\cite{Halperin2011} (see Supp. Section XIV). For the smallest interferometer, we obtain a charging energy of $18\,\mu$eV similar to that of GaAs devices of similar sizes operating in the Aharonov-Bohm regime~\cite{Nakamura2019}, and $\xi \simeq 6\times 10^{-3}$, which is fully consistent with the absence of Coulomb blockade in our graphene devices.  Yet, we also observed Aharonov-Bohm interference in two other graphene devices equipped instead with a $285$~nm thick SiO$_2$ back-gate dielectric (see Supp. Section XIII). Despite a reduction of about 15 of the back-gate surface capacitance, we estimate $\xi \simeq 7\times 10^{-2}$ which remains consistent with the Aharanov-Bohm regime. The systematic presence of back-gate electrodes, and to a lesser extent, the close proximity of the top-gate electrodes in hBN-encapsulated graphene therefore provide efficient screening of charging effects that enables to observe Aharonov-Bohm interference.

The magnetic-field period $\Delta B$ of the Aharanov-Bohm oscillations provides a direct measure of the effective area $A_{\text{AB}}$ drawn by the interfering edge channel. For the three interferometers we obtained $\Delta B=1.32$, $0.40$ and $0.27$ mT corresponding to $A_{AB} = 3.1$, $10.4$ and $15.0\,\mu\text{m}^2$, which is consistent with our expected geometrical areas of $3.1\pm0.4$, $10.7\pm1.2$ and $14.7\pm1.8$ $\mu\text{m}^2$, substantiating the $\phi_0$-periodicity obtained with the electrostatic analysis of the plunger-gate effect. Notice that the precision of the geometrical areas is limited by the uncertainty in the optical determination of the graphene physical edges position, which could easily explain the small differences with the measured Aharonov-Bohm areas. \\

\textbf{Decoherence and thermal broadening}

The loss of visibility in QH interferometers is a fundamental question that encompasses several phenomena such as thermal broadening or quantum decoherence by inelastic processes and energy relaxation.
We investigated the coherence properties of our interferometers through the joint analysis of the bias and temperature dependence of the visibility. The multiple QPCs configuration enables us to systematically study the coherence properties for three different cavity lengths. We begin with the out-of-equilibrium measurements performed with a dc voltage bias $V_{\text{dc}}$ applied on the source contact (the drain contact is kept grounded). Electrons injected at a finite energy $\delta \epsilon$ above the Fermi energy of the cavity have a phase shift $2\pi \delta \epsilon 2L/hv$ proportional to twice the length $L$ of the edge channel between the two QPCs and to the edge-excitation velocity $v$. This additional phase shift that adds up to the Aharonov-Bohm phase can yield theoretically a variety of oscillation patterns as a function of magnetic flux and voltage bias, which depend on the voltage drop across the device (see Supp. Section XV). For a symmetric drop at the two QPCs ($V_{\text{dc}}$/2 and $-V_{\text{dc}}$/2 on the source and drain), the resistance oscillations draw a checkerboard pattern of the form $\delta R_{\text{D}}\propto
\text{cos}(2\pi\phi/\phi_{0})\text{cos}(2 \pi e V_{\text{dc}}/E_{\text{Th}})$. For a fully asymmetric drop ($V_{\text{dc}}$ and $0$ on the source and drain) a diagonal strip pattern of the form $\delta R_{\text{D}}\propto \text{cos}(2\pi\phi/\phi_{0} - 4\pi e V_{\text{dc}}/E_{\text{Th}})$ is expected instead. In these expressions, the oscillations period as a function of $V_{\text{dc}}$ is governed by the ballistic Thouless energy $E_{\text{Th}} = h/\tau$ related to the traveling time $\tau=L/v$ between the two QPCs~\cite{Chamon1997,NgoDinh2012}. 

For the large interferometer at $B=14$~T, the resistance oscillations at finite bias (Fig.~3d) draw a checkerboard pattern similar to those observed in GaAs devices~\cite{McClure2009,Yamauchi2009,Nakamura2019}. The resulting oscillations decay quickly as a function of the dc voltage $V^{\text{dc}}_{\text{D}}$ (measured across the diagonal contacts $V_{\text{D}}^+$ and $V_{\text{D}}^-$), indicating that some energy relaxation processes are at play at finite bias (see Supp. Fig.~S18). Repeating the measurements on the two other interferometers yields similar patterns (see Supp. Section XV), though with a notable difference in the form of the checkerboard which tends to be tilted towards a diagonal strip pattern upon reducing the interferometer size, as illustrated in Fig.~3c for the small interferometer. This tilt can be accounted for by an incomplete equilibration of chemical potential carried by the electron flow, which partially maintains the asymmetric voltage drop across the cavity. Calculation of the theoretical FP transmission with an unbalanced voltage drop describes very well the observed tilted checkerboard, as shown in Fig.~3e and f (theoretical analysis and additional data are provided in Supp. Section XV).

The key parameter extracted from these checkerboards is the Thouless energy of the cavity $E_{\text{Th}}=hv/L$ that is given by the $V^{\text{dc}}_{\text{D}}$-periodicity. The resulting values shown in Fig.~3i for the three interferometers are found to consistently scale with $1/L$. 

Besides, theory predicts that the temperature dependence of the oscillations visibility also relates to the Thouless energy~\cite{Chamon1997}. The blurring of interference by the thermal broadening of the impinging electrons leads to an exponential suppression of the visibility with temperature, which follows $\exp(-4\pi^2 k_\text{B} T/E_{\text{Th}})$ with $k_\text{B}$ the Boltzmann constant. In Fig.~3g and h we show the $T$-dependence of the resistance oscillations for the three interferometers. A clear exponential suppression of the visibilities is obtained in all cases, which is fitted with $\exp(-T/T_0)$ (dashed lines in Fig.~3h). The resulting $4\pi^2 k_\text{B} T_0$ values that we append to Fig.~3i conspicuously scale as $1/L$ and are furthermore in excellent agreement with the Thouless energies extracted from the checkerboard patterns, bearing out a visibility limited mainly by thermal broadening.  The slope in Fig.~3i further enables us to extract an estimate of the edge velocity $v  = 1.4\times10^{5}$~m.s$^{-1}$, which is of the same order as that obtained in GaAs devices~\cite{McClure2009,Gurman16,Nakamura2019}, though here at a much higher magnetic field. Finally, a phase coherence length of  $10\,\mu$m at our base temperature can be assessed from the exponential suppression of the visibility with the perimeter of the interferometer, after corrections accounting for thermal broadening (see Fig. S19). 
This global set of data that complies with most of the theoretical expectations~\cite{Chamon1997} therefore demonstrates graphene to be a highly tunable, model platform for QH-FP interferometry. \\

\textbf{Coherently-coupled double FP interferometer}  

Here we operate our device in a three QPCs configuration to show that the large FP interferometer remains coherent upon turning on backscattering at the middle QPC. As sketched in Fig.~4a and b, Aharonov-Bohm interference depends upon three fluxes, $\phi_{1,2,3}$, defined by the three cavity areas (blue, green and yellow in Fig.~4b, respectively). Each of them can be tuned by the magnetic field and one or two plunger gates. Figure~4c  displays the resistance of the device for the three QPCs tuned at partial transmission of the outer edge channel, with a total transmission $T\simeq 0.46$, upon varying both plunger-gate voltages $V_{\text{pg1}}$ and  $V_{\text{pg2}}$. The resistance oscillates with both gate voltages and draws a regular pattern characteristic of the flux variation $\phi_1$ with $V_{\text{pg1}}$  (blue FP cavity), and $\phi_2$ with $V_{\text{pg2}}$ (green  FP cavity). Coherence through the double interferometer is unveiled in the 2D Fourier transform in Fig.~4e. In addition to the two peaks at frequencies $(f_{\text{pg1}} , f_{\text{pg2}})=(53\,\text{V}^{-1},0\,\text{V}^{-1})$ and $(0\,\text{V}^{-1},60\,\text{V}^{-1})$ corresponding to fluxes $\phi_1$ and $\phi_2$ respectively, a third peak emerges at $(f_{\text{pg1}} , f_{\text{pg2}})=(53\,\text{V}^{-1},60\,\text{V}^{-1})$, indicative of a joint modulation by both plunger gates and hence a modulation of the double cavity flux $\phi_3$. Inspection of the four quadrants of the Fourier transform in Fig. S21 shows that a fourth peak related to a $\phi_{1}-\phi_{2}$ contribution is present but with a lower amplitude than the $\phi_3$ contribution. This indicates that the $\phi_3$ peak is mainly the result of the interference process for which electron wavefunctions interfere coherently after passing twice (back and forth) through the partially-transmitting middle QPC. (see Supp. Section XVIII for a detailed theoretical analysis)

Figure~4d shows a complementary measurement where the magnetic field is varied together with the plunger-gate voltage $V_{\text{pg2}}$ acting on the area of the small interferometer. We observe diagonal stripes similar to those in Fig.~3a and characteristic of the Aharonov-Bohm effect for the small cavity with flux $\phi_2$ (see peak at $(f_{B} , f_{\text{pg2}})=(0.79\,\text{mT}^{-1},90\,\text{V}^{-1})$ in Fig.~4f), but with an additional wiggling. This wiggling results from the Aharonov-Bohm oscillations of the medium interferometer (in series with the small one) via the flux $\phi_1$, which is independent of $V_{\text{pg2}}$ and corresponds to the peak at  $(f_{B} , f_{\text{pg2}})=(2.54\,\text{mT}^{-1},0\,\text{V}^{-1})$ in the Fourier transform (see Fig.~4f). The double cavity flux $\phi_3$, which depends upon both $B$ and $V_{\text{pg2}}$, leads to a peak at $(f_{B} , f_{\text{pg2}})=(3.49\,\text{mT}^{-1},90\,\text{V}^{-1})$ with a magnetic field periodicity which corresponds to the double cavity area, that is, the sum of the small and medium cavity areas. The precise shape of the wiggling of the diagonal stripes is a direct evidence of the contribution of the double cavity flux (see Supp. Section XVIII for an additional measurement configuration).

The observation of the flux periodicity $\phi_3$, together with a careful Fourier analysis in the four quadrants (see Supp. Section XVIII), provides compelling evidence that electron interference occurs through the central QPC at partial transmission, and hence that both FP cavities are coherently coupled. Such a tunable multiple FP interferometer paves the way to more advanced devices in which sequential transfer of single particles at the central QPC, controlled for instance with an anti-dot in the QPC constriction, would enable braiding schemes for non-Abelian anyons~\cite{DasSarma05}. \\

\textbf{Conclusion and outlook}  

Finally, the high-visibility Aharonov-Bohm interference that we observe with remarkable agreement with the non-interacting theory~\cite{Chamon1997} demonstrates the relevance of graphene for performing prototypical QH-FP interferometry with integer quantum Hall edge channels. The high mobility and the versatility of the graphene van der Waals hetereostructures turn out to be pivotal to harness fine control of QH edge-channel transmissions in QPCs~\cite{Zimmermann2017}, and therefore construct advanced gate-tunable interferometers. With further study in the fractional quantum Hall regime, this graphene platform gives new opportunities for anyon physics in QH interferometers, potentially extendable to time-resolved electron quantum optics experiments~\cite{bauerle18}. Besides, the recent advances in coupling graphene QH edge channels with superconductivity~\cite{Amet16,Lee17,Zhao20} may lead to a variety of novel interferometry devices~\cite{Huang19} in which proximity-induced topological superconductivity could be intertwined with QH interferometry for readout or braiding schemes. Such perspectives could in turn open alternative pathways for quantum-information processing of topological excitations~\cite{Clark2013,Stern13,Mong14}.

\textit{Note}: A very recent work (https://arxiv.org/abs/2008.12285) that appeared during the reviewing of our manuscript confirmed our conclusions with similar graphene devices.

\section*{Methods}
	
\subsection*{Sample fabrication}
hBN/graphene/hBN heterostructures were assembled from exfoliated flakes using the van der Waals pick-up technique~\cite{Wang13}. The substrates are highly doped Si wafers with a $285$~nm thick SiO$_2$ layer. For the sample discussed in the main text (BNGr74), the heterostructure is deposited on a thin layer of graphite that serves as a back-gate electrode.  Contacts were patterned by electron-beam lithography and metalized by e-gun evaporation of a Cr/Au bilayer after etching of the stack with a CHF$_3$/O$_2$ plasma directly through the resist pattern used to define the contacts. The electrostatic plunger and split gates were obtained with a second electron-beam lithography step and subsequent evaporation of Pd. The graphite layer for sample BNGr74 was also contacted at this step on a purposely uncovered part. Two other samples (BNGr64 and BNGr30) discussed in Supp. Section XIII were prepared without graphite-gate electrode but with the hBN/graphene/hBN heterostructure resting directly atop the Si/SiO$_2$ substrate. 

\subsection*{Measurements}
Measurements were performed in a dilution fridge reaching a base temperature of $0.01$~K and equipped with a superconducting solenoid. To ensure good electron thermalization, the fridge is equipped with room-temperature feedthrough filters, highly-dissipative wiring, copper-powder filters at the mixing chamber stage, and cryogenic-compatible capacitors to ground on each line mounted directly on the sample holder. Devices were measured in four-terminal, voltage-bias configuration using an ac voltage of 5 $\mu$V and low-frequency lock-in amplifier techniques. Current was measured with a home-made current-voltage converter. Non-linear transport measurements were carried out by adding a dc voltage between source and drain contacts and measuring the dc and ac components of the diagonal voltage $V_{\rm D}$. Measurements of the resistance oscillations as a function of magnetic field and plunger-gate voltage were performed by using the current decay of the superconducting solenoid in persistent mode while sweeping the plunger-gate voltage with a 20 bits home-made voltage source. All room-temperature low-noise pre-amplifiers were thermalized in a home-made, temperature-controlled box to get rid of thermal drifts of input voltage offsets. 

\section*{Data availability}
The data that support the findings of this study are available from the corresponding author upon reasonable request.

\section*{Acknowledgments}
We thank I. Aleiner for valuable discussions. We thank S. Dumont for the development of dedicated low-noise, high stability voltage sources and F. Blondelle for his technical support.
Samples were prepared at the Nanofab facility of N\'{e}el Institute. 
This work was supported by the H2020 ERC grant \textit{QUEST} No. 637815. 
K.W. and T.T. acknowledge support from the Elemental Strategy Initiative conducted by the MEXT, Japan, Grant Number JPMXP0112101001, JSPS KAKENHI Grant Number JP20H00354 and the CREST(JPMJCR15F3), JST.

\section*{Competing Interests} The authors declare that they have no competing financial interests.

\section*{Author contributions}
C.D., L.V., H.V. and G.N. made the sample fabrication. C.D. performed the experiments under the supervision of B.S.. F.G. provided technical support on the experiments. K.W. and T.T. grew the hBN crystals. C.D., H.S. and B.S. analyzed the data. C.D. and H.S. made the theoretical developments. B.S. conceived the project and wrote the paper with the inputs of all co-authors.

\begin{figure*}[h!]
\includegraphics[width=0.9\linewidth]{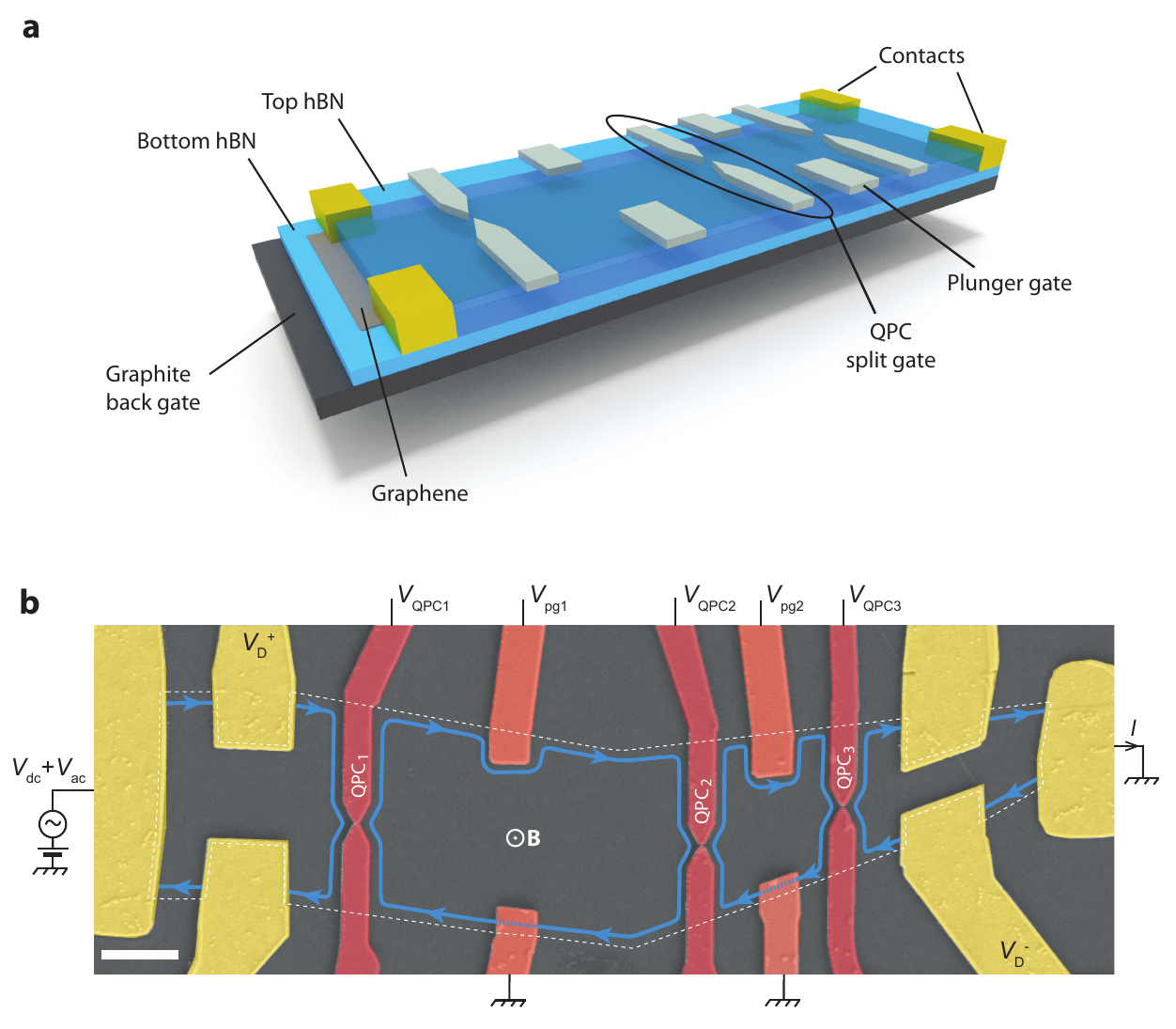}
\centering
\caption{\textbf{Graphene quantum Hall Fabry-P\'{e}rot interferometer.} \textbf{a,} Schematic of the van der Waals hBN/graphene/hBN/graphite heterostructure equipped with split-gate and plunger-gate electrodes (light gray) atop it. The graphite flake serves as back-gate electrode and the graphene is contacted through edge contacts~\cite{Wang13} (yellow). Further details about nanofabrication are given in Methods. \textbf{b,} False-colored scanning electron micrograph of the device. The scale bar is  1 $\mu$m. Three QPCs define two FP cavities. The interfering quantum Hall edge channel (blue line) propagates along the graphene edges (white dashed line), and along the split-gate (red) and plunger-gate (orange) electrodes, illustrating a configuration in which the gate electrodes deplete the charge carriers and repel the quantum Hall edge channel. The transmissions of the FP cavities are measured through the diagonal differential resistance $R_{\text{D}}=\text{d}V_{\text{D}}/\text{d}I$, where $I$ is the current measured by an ampmeter and $V_{\text{D}} $ the diagonal voltage drop across the contacts $V_{\text{D}}^+$ and $V_{\text{D}}^-$, in voltage bias configuration using a dc and ac voltage sources, $V_{\text{dc}}$ and $V_{\text{ac}}$.}
\label{Fig1}
\end{figure*}

\begin{figure*}[h!]
\centering
	\includegraphics[width=1\linewidth]{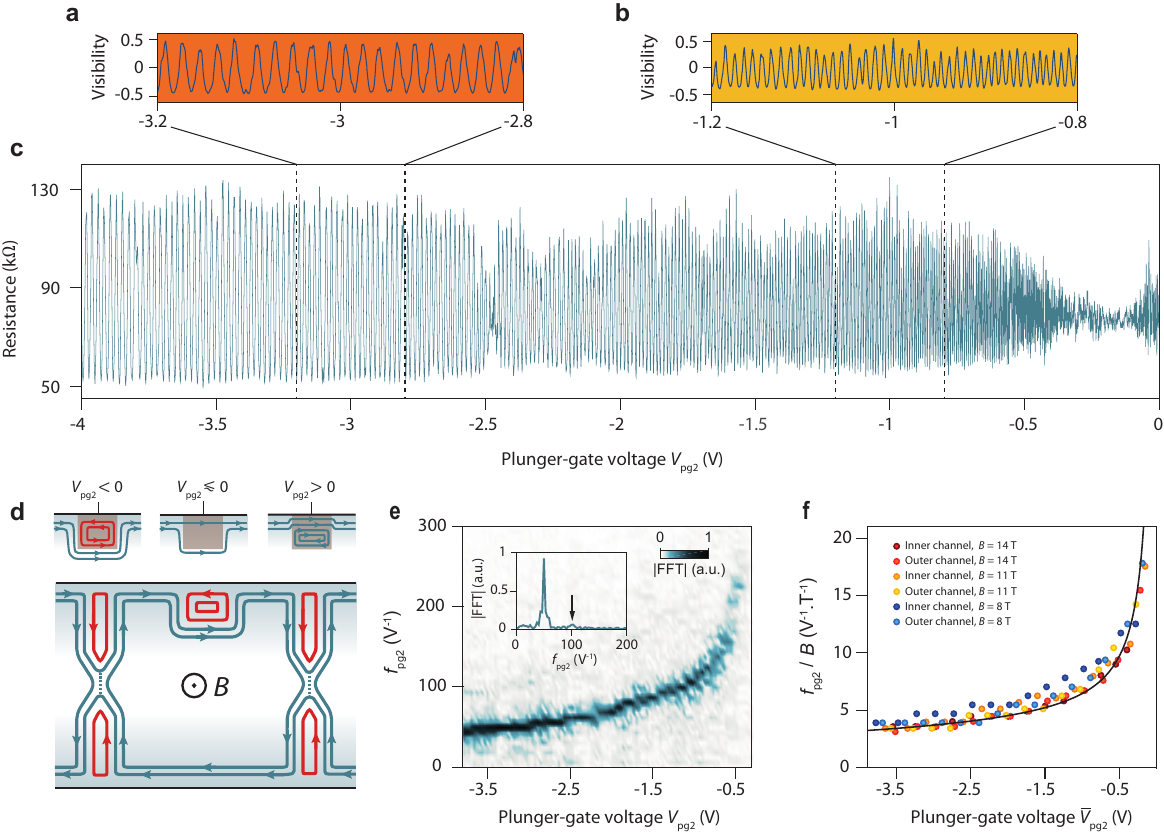}
	\caption{\textbf{Gate-tunable quantum interference.} 
	\textbf{a, b, c,} Diagonal resistance oscillations as a function of plunger-gate voltage $V_{\text{pg2}}$ measured on the small interferometer at $0.015$~K and $14$~T (with an ac bias voltage of $5\,\mu$V). The back-gate voltage is set to $V_{\text{bg}}=0.533$~V corresponding to a filling factor $\nu_b=1.5$ in the bulk. The split-gate voltages on QPC$_2$ and QPC$_3$ are tuned to obtain partial transmission of the outer edge channel. The charge neutrality point below the plunger gate is at $-0.3$~V and corresponds to a suppression of the oscillation amplitude in c due to the divergence of the oscillation frequency shown in e. \textbf{a} and \textbf{b} show zooms on smaller $V_{\text{pg2}}$ ranges of the resistance oscillations converted in visibility $(R-\bar{R})/\bar{R}$, where  $\bar R$ is the resistance average. 
\textbf{d,} Schematics of the QH-FP interferometer illustrating the edge channels configuration for the measurements in c. The black lines represent the physical edges of graphene.  The blue (red) lines indicate electron(hole)-like edge channels and the arrows the direction of motion of charge carriers. At the QPC constriction, the dashed line indicates the tunneling of the interfering edge channel. Top sketches: Three configurations for the states around the plunger gate. Left, accumulation of localized hole states repelling the propagating electron edge channels. Middle, depletion of charge carrier density to a filling factor 1 below the gate, which repels the inner edge channel. Right, accumulation of localized electron states that push the propagating edge channels closer to the graphene edge. 
\textbf{e,} Fourier amplitude of the resistance oscillations in c as a function of $V_{\text{pg2}}$ and the plunger-gate-voltage frequency $f_{\text{pg2}}$ obtained by computing the Fourier transform over a small $V_{\text{pg2}}$ window of $0.16$~V  sliding over the whole $V_{\text{pg2}}$ range.
  Inset : Fourier transform at $V_{\text{pg2}}= -3.28$~V  showing a well-defined peak at $f_{\text{pg2}}$ = 50 $\text{V}^{-1}$ and a faint peak at $f_{\text{pg2}}$ = 100 $\text{V}^{-1}$ indicated by the black arrow. These peaks correspond to first order and second order (two turns in the FP loop) interference processes. 
\textbf{f,} Evolution of the main peak frequency $f_{\text{pg2}}$ rescaled by the magnetic field $B$ as a function of $\bar{V}_{\text{pg2}}$, the plunger-gate voltage shifted with respect to the voltage that expels the interfering edge channel. The plot gathers a set of experiments performed with different interfering edge states and at different magnetic fields. The collapse of all data points into a single curve is fitted by an electrostatic simulation  (see Supp. Section IX) of the pn-junction displacement with plunger-gate voltage (black line).}
		\label{Fig2}
\end{figure*}

 \begin{figure*}[h!]
 \centering
 	\includegraphics[width=1\linewidth]{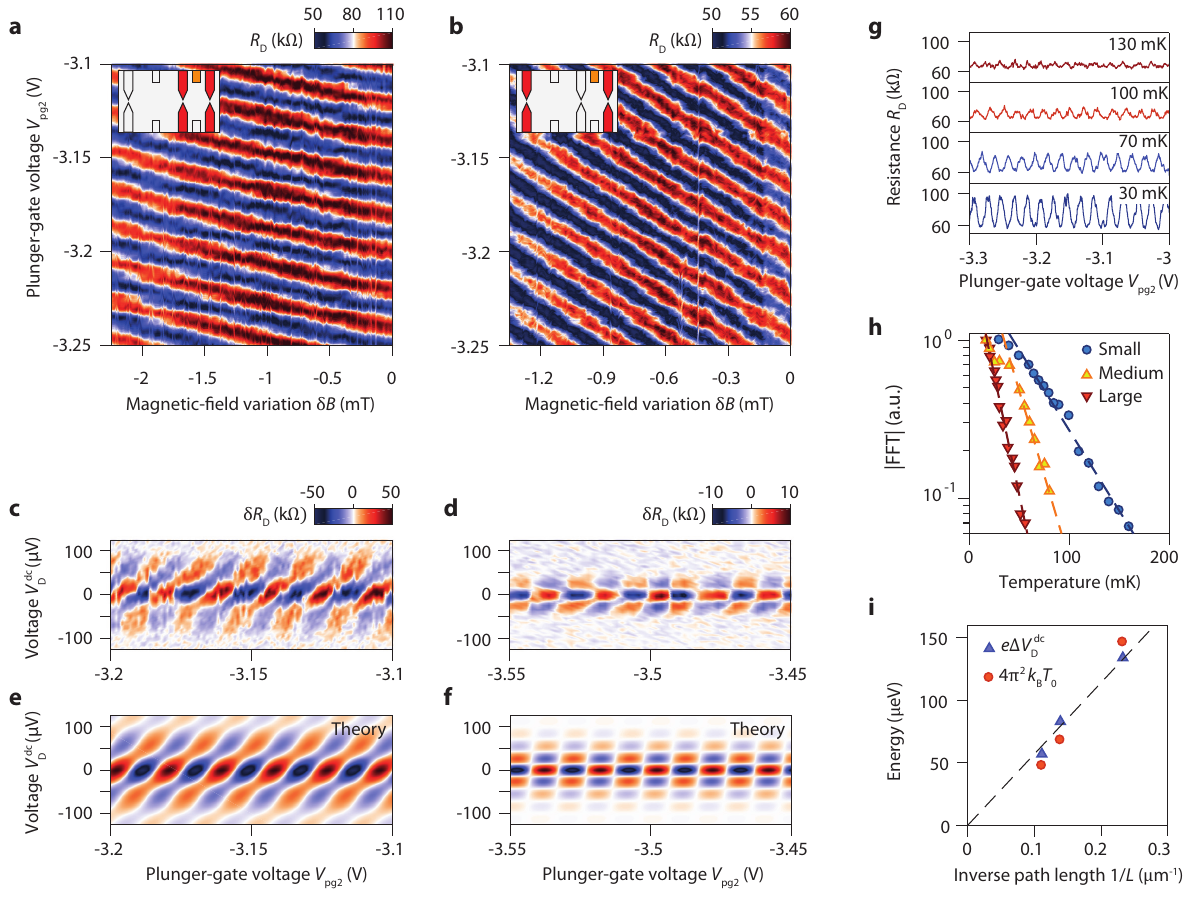}
 	\label{Fig3}
 	 	\caption{\textbf{Aharonov-Bohm effect and energy dependence.} 
	\textbf{a, b,} Diagonal resistance $R_{\text{D}}$ versus plunger-gate voltage $V_{\text{pg2}}$ and magnetic field variation $\delta B$ measured at $0.015$~K for the small interferometer in a and the large interferometer in b. The inset schematics indicate the active QPCs (in red) and plunger gates (in orange) for the respective measurements. The negative slope in the $B-V_{\text{pg2}}$ plane is a clear indicator for the Aharonov-Bohm effect in both interferometers.
\textbf{c, d,} Differential diagonal resistance variations $\delta R_D$, after background subtraction, versus measured dc diagonal voltage $V_{\text{D}}^{\text{dc}}$ and plunger-gate voltage $V_{\text{pg2}}$ for the small interferometer in c and the large interferometer in d.
Typical checkerboard patterns are observed with a significant tilt for the smallest interferometer revealing incomplete symmetrization of the voltage bias. 
\textbf{e, f,} Numerical simulations of resistance oscillations induced by voltage bias and plunger-gate voltage that reproduce the data presented in c and d, respectively. The simulation incorporates an asymmetric potential drop at the two QPCs and an out-of-equilibrium decoherence factor (see Supp. Section XV). The asymmetry factor is $x=0.2$ in e and $x=0.02$ in f. 
\textbf{g,} Temperature evolution of the resistance oscillations versus plunger-gate voltage $V_{\text{pg2}}$ for the small interferometer. \textbf{h,} Exponential decays of the Fourier amplitude of the resistance oscillations for the small (blue), medium (yellow) and large (red) interferometers. The dashed line is a fit  with $\text{exp}(-T/T_0)$ giving $T_0 = 43$, $20$, and $14$ mK for the small, medium, and large interferometers, respectively. \textbf{i,} Evolution of the  energy scales as a function of the inverse of the cavity length $L$ of the three interferometers. The oscillation period with the dc voltage, $\Delta V_{\text{D}}^{\text{dc}}$, of the checkerboard pattern corresponds to the Thouless energy $E_{\text{Th}}$ and is expected to be equal to the energy scale $4\pi^2 k_{\rm B\it} T_0$ of the temperature-induced blurring of the resistance oscillations~\cite{Chamon1997}. The dashed line is a linear fit highlighting the $1/L$ scaling of both energy scales. All data in this figure are obtained in a configuration with the outer edge channel interfering at $B=14$~T.}
 	 \end{figure*}

\begin{figure*}[h!]
\centering
	\includegraphics[width=1\linewidth]{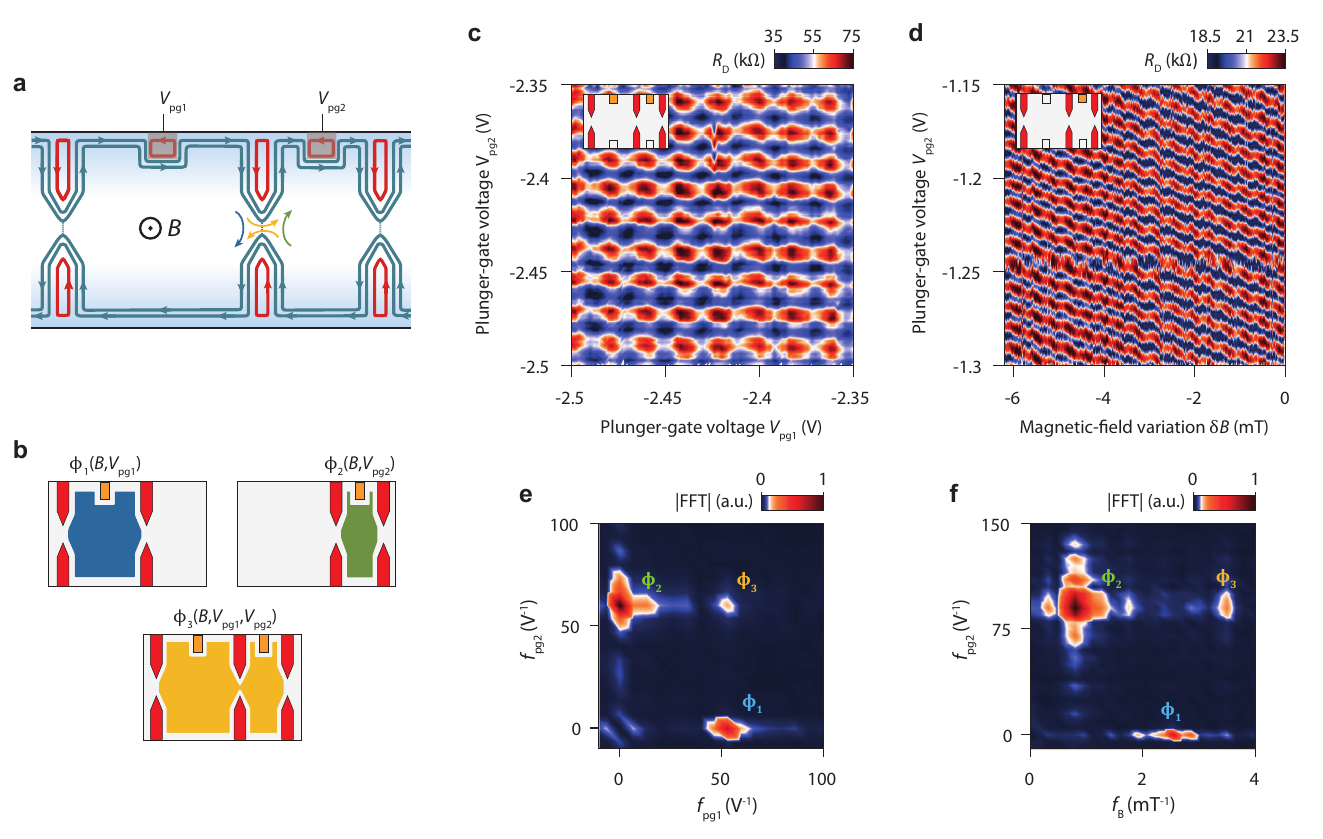}
	\label{Fig4}
	\caption{\textbf{Coherently-coupled double QH-FP interferometer.} 
	\textbf{a,} Schematics of the double QH-FP interferometer in a configuration of partial backscattering of the inner edge channel at the three QPCs. The central QPC can either reflect an incoming electron from the left cavity (blue arrow) or the right cavity (green arrow), or transmit coherently an electron from one cavity to the other (yellow arrows). In the latter case, electrons pick up the Aharonov-Bohm phase given by the combined areas of the small and medium interferometers. 
	\textbf{b,}  Schematics of the cavities involved in the different interference processes depicted in a. $\phi_{1,2,3}$ are the corresponding Aharonov-Bohm fluxes, indicated with the magnetic field and plunger-gate voltages that modulate them. \textbf{c,} Diagonal resistance versus plunger-gate voltages $V_{\text{pg1}}$ and $V_{\text{pg2}}$ (outer edge channel interfering, $B=14$~T). 
	\textbf{d,} Diagonal resistance $R_{\text{D}}$ versus magnetic field variation $\delta B$ and plunger-gate voltage $V_{\text{pg2}}$ (inner edge channel interfering, $B=14$~T). The inset schematics in c and d indicate the active QPCs (in red) and plunger gates (in orange) for the respective measurements. 
	\textbf{e,} Fourier amplitude of the resistance oscillations in c versus plunger-gate-voltage frequencies $f_{\text{pg1}}$ and $f_{\text{pg2}}$. The three peaks at $(f_{\text{pg1}} , f_{\text{pg2}})=(53\,\text{V}^{-1},0\,\text{V}^{-1})$, $(0\,\text{V}^{-1},60\,\text{V}^{-1})$  and $(53\,\text{V}^{-1},60\,\text{V}^{-1}) $ correspond to the three Aharonov-Bohm fluxes depicted in b. 
	\textbf{f, } Fourier amplitude of the resistance oscillations in d versus magnetic field frequency $f_{\text{B}}$ and plunger-gate-voltage frequency $f_{\text{pg2}}$. The three peaks of the three Aharonov-Bohm fluxes emerge at $(f_{B} , f_{\text{pg2}})=(0.79\,\text{mT}^{-1},90\,\text{V}^{-1})$, $(2.54\,\text{mT}^{-1},0\,\text{V}^{-1})$ and $(3.49\,\text{mT}^{-1},90\,\text{V}^{-1})$. In both Fourier amplitude maps, the peak that corresponds to flux $\phi_3(B,V_{\text{pg1}},V_{\text{pg2}})$ reveals the quantum coherence throughout the two cavities of the double QH-FP interferometer.}

\end{figure*}

\bibliography{QH-FPI-BIB}

\clearpage


\clearpage
\onecolumngrid
\setcounter{figure}{0}
\setcounter{section}{0}
\renewcommand{\thefigure}{S\arabic{figure}}

\newpage

\part*{ \centering Supplementary Information}
\bigskip 


	\section{Samples studied}

\begin{figure}[h!]
	\centering
	\includegraphics[width=1\textwidth]{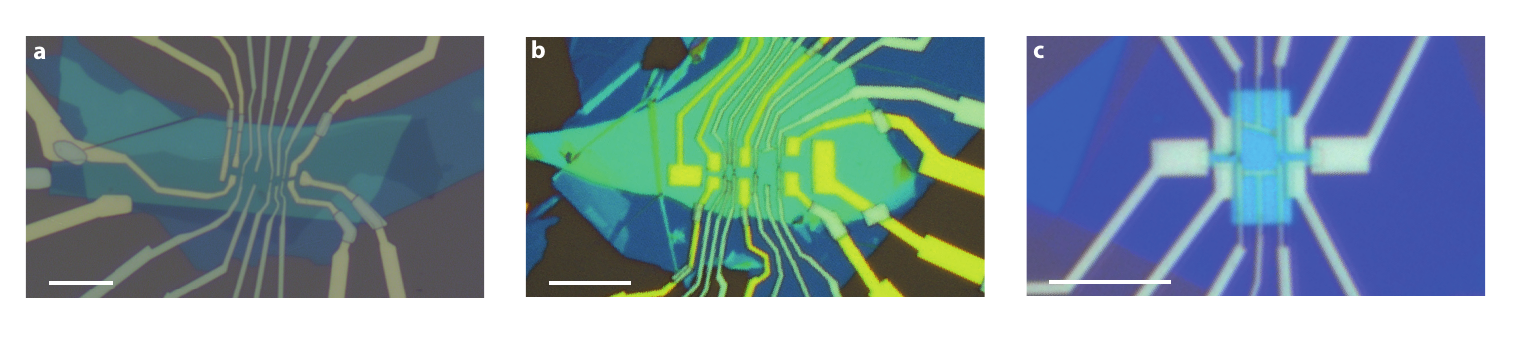}
	\caption{\textbf{Optical images of the devices.} \textbf{a,} Sample BNGr74 of the main text. \textbf{b,} Sample BNGr64 described in section \ref{OtherDevices}. \textbf{c,} Sample BNGr30 described in section \ref{OtherDevices}. Scale bars are 10 $\mu\rm m$.}
	\label{fig:Image_samples}
\end{figure}

Figure~\ref{fig:Image_samples} displays optical images of the devices studied in this work. The fabrication process is described in Methods. The thickness of the van der Waals layers and the size of the split-gate gaps are reported in Table~\ref{Table1}.

{\small
\begin{table*}[h!]

\centering
\begin{tabular}{|c|c|c|c|c|}
\hline
  Sample  & Top hBN & Bottom hBN  & Graphite  & Split-gate \\
  & thickness (nm) & thickness (nm)  & thickness (nm) & gap (nm)
  \\
  \hline \hline
  BNGr74 &  22 & 18 & $4$ &  0 / 21 / 20 \\
  \hline
  BNGr64 & 20 & 50 &  &   148 / 159\\
  \hline
  BNGr30 & 25 & 15  & & 129 / 140\\
  \hline   
\end{tabular}
\caption{\textbf{Samples characteristics.} The thickness of the hBN and graphite layers are measured by atomic force microscopy. The gap size of the split-gate electrodes is measured by scanning electron microscopy.}
 \label{Table1}
\end{table*}
}

\section{Parameters extracted from the Aharonov-Bohm interference}
	
Table~\ref{Table2} presents the various parameters extracted from the measurements shown in Figure 3 of the main text, among which the Aharonov-Bohm period $\Delta B_{@14\,\rm T}$, the Thouless energy $E_{\text{Th}}$ extracted from the checkerboard patterns, the energy scale $T_0$ related to the temperature dependence of the resistance oscillation, together with the geometrical dimensions (surfaces and lengths between QPCs) of the three Fabry-P\'{e}rot cavities.

Importantly, we stress that the determination of the device geometry relies on optical images of the graphene flake taken during the van der Waals pick up process, which makes the exact determination of the graphene edge delicate. We therefore assess the graphene edge position from this image at $\pm 150$~nm, which results in the uncertainties of the geometrical area $A_{\rm geo}$ of the FP cavities and lengths $L$ between QPCs reported in Table~\ref{Table2}.

\begin{table*}[h!]
\centering
\begin{tabular}{|l|c|c|c|c|c|c|c|}
\hline
 QH-FP  & $\Delta B_{@14\,\rm T}$ & $A_{\text{AB}}$& $A_{\text{geo}}$ & $L$ & $E_{\text{Th}}$ & $T_0$& $E_{\text{Th}}/4\pi^2k_{\text{B}} $
 \\
 & (mT) &($\mu \text{m}^2$)&$(\mu \text{m}^2)$  & ($\mu \text{m}$) & ($\mu V$)& (mK)&(mK)\\
  \hline \hline
 Small & $1.32$ &  $3.1$& $3.1\pm0.4$ & $4.3\pm0.5$& 134  & 43 & 39\\
 \hline
  Medium  & $0.40$ & $10.4$&$10.7\pm1.2$ & $7.2\pm0.5$ & 83  & 20 & 24\\
  \hline
  Large & $0.27$ &   $15.0$& $14.7\pm1.8$ & $9.0\pm0.5$ & 57 & 14 & 17\\
  \hline   
\end{tabular}
    \caption{\textbf{Geometrical and physical parameters corresponding to the measurements of Fig. 3} Aharonov-Bohm period $\Delta B_{@14\,\rm T}$ obtained at $B=14$~T and resulting Aharonov-Bohm area $A_{\text{AB}}$;   geometrical area $A_{\rm geo}$ of the FP cavities; geometrical length $L$ between two QPCs of the cavity; Thouless energy $E_{\text{Th}}$ extracted from the checkerboard patterns in Fig.~3c and d; Energy scale $T_0$ extracted from Fig.~3h; $E_{\text{Th}}/4\pi^2 k_{\text{B}}$, the quantity theoretically equal to $T_0$ according to ref.~\cite{Chamon1997}}
    \label{Table2}
\end{table*}

\newpage
\clearpage
\section{Design characteristic of the QPCs}

\begin{figure}[h!]
  	\centering
  	\includegraphics[width=1\textwidth]{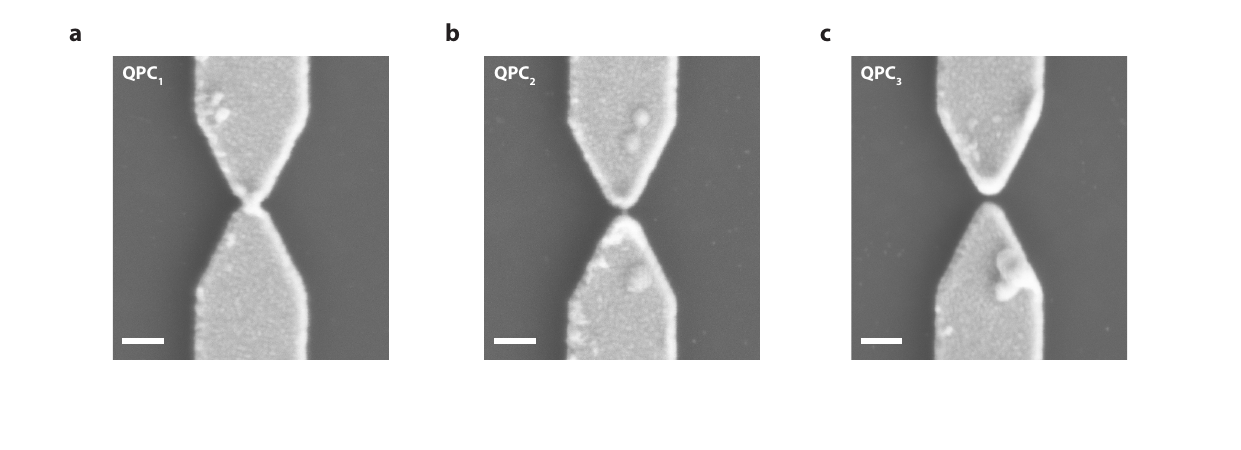}
  	\caption{\textbf{QPCs geometry.} Scanning electron micrograph of the QPCs of BNGr74 device. \textbf{a,} QPC$_1$. \textbf{b,} QPC$_2$. \textbf{c,} QPC$_3$. The two split-gate electrodes of QPC$_1$ are unintentionally connected. The gaps between the two other split gates is 20 nm. Scale bar is $100$ nm.}
	\label{fig:Image_QPC}		
  \end{figure}

The presence of the graphite back-gate electrode separated from the graphene by a thin hBN dielectric layer imposes drastic conditions for the design of the split-gate electrodes. Contrary to devices on Si/SiO$_2$ studied in ref.~\cite{Zimmermann2017,Veyrat19} in which the split-gate gap of about $150$~nm led to a suitable ratio of split-gate capacitance to QPC capacitance, the very close proximity of the graphite back gate imposes a much smaller split-gate gap. By performing numerical simulations~\cite{Veyrat19}, we estimated the split-gate gap 
that leads to a ratio of split-gate capacitance to QPC capacitance of the order of 2 to be of the order of few tens of nanometers, depending on the hBN thicknesses. Figure~\ref{fig:Image_QPC} displays scanning electron micrographs of the three split gates of sample BNGr74 discussed in the main text. The split-gate gaps of QPC$_2$ and QPC$_3$ are $20$~nm, suitable for operating the split gates as QPCs in the quantum Hall regime. The split-gate electrodes of QPC$_1$ are unintentionally connected but this short-circuit does not hinder QPC operation (see QPC characterizations in section \ref{section_QPC_14T}).

\newpage
\clearpage
\section{Characterization of the split-gate capacitances at $\boldsymbol{0}$~T}\label{section_QPC_0T}

\begin{figure}[h!]
	\centering
	\includegraphics[width=\textwidth]{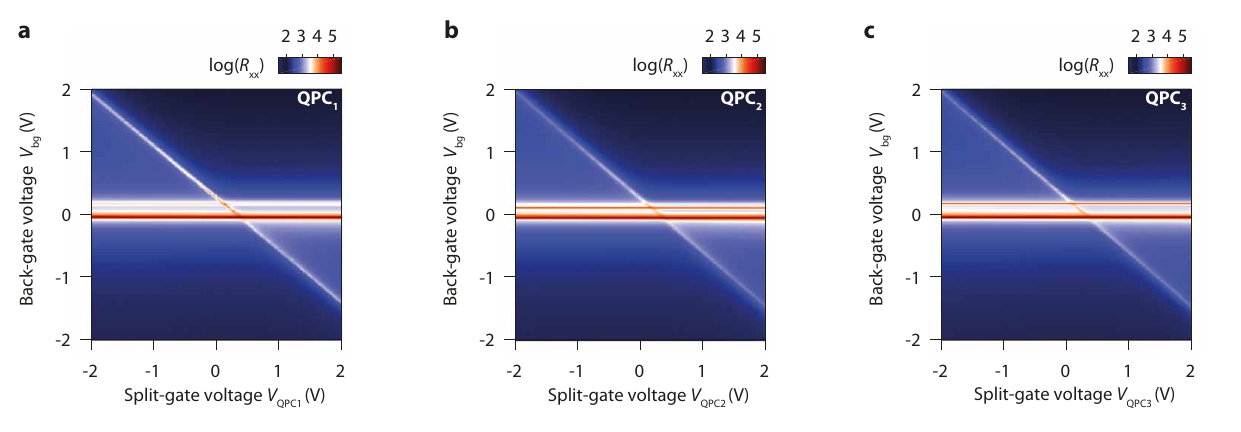}
	\caption{\textbf{Split-gates characterization at $\boldsymbol{0}~\text{T}$.} \textbf{a, b, c,} Longitudinal resistance $\it R \rm_{\text{xx}}$ versus split-gate voltage $\it V\rm_{QPC}$ and back-gate voltage $\it V\rm_{bg}$ for the three QPCs of the QH-FP interferometer presented in the main text. The horizontal line at $\it V\rm_{bg} = - 0.04$~V corresponds to the charge neutrality point in the bulk graphene, whereas the diagonal lines correspond to the charge neutrality point in the graphene beneath the active split gate. These lines intersect at ($\it V\rm_{QPC},\it V\rm_{bg}) \simeq (+0.38~V, -0.04 ~V)$ revealing the significant local hole doping induced by the palladium gates. The second horizontal line at $\it V\rm_{bg} = 0.18$~ V in a and c and  $\it V\rm_{bg} = 0.12$~V in b marks the positive back-gate voltage needed to compensate the hole doping induced by the palladium beneath the non-active split gates.}
	\label{fig:QPC_0T}
\end{figure}

In this section, we present the characterization of the back gate and the different split gates at zero magnetic field for the sample of the main text. Figure~\ref{fig:QPC_0T} shows color-coded maps of the longitudinal resistance $R_{\rm xx}$ versus back-gate voltage $V_{\rm bg}$ and voltage $V_{\rm QPC}$ applied on a split gate (other split gates are floating). 
The maps exhibit four quadrants separated by two nearby horizontal lines and a diagonal line. The most resistive horizontal line, at $V_{\rm bg}=-0.04~\rm V$, corresponds to the charge neutrality point in the bulk of graphene and the diagonal line corresponds to the charge neutrality point below the active split-gate electrodes, as usual for graphene devices equipped with a local top gate. The two lines intersect at $V_{\rm QPC}\simeq +0.38$~V as a result of the work function difference between the palladium of the gates and the graphene. The second horizontal line is more unusual and results from the local hole doping of the graphene beneath the two other split gates that are not active but contribute in series to the measured resistance. The palladium of these split-gate electrodes shifts the position of the charge neutrality point beneath them to $V_{\rm bg}=0.12-0.18~\rm V$, yielding a secondary resistance peak, independent of the active split gate and observed consistently for the three QPC maps. These maps also provide the capacitance ratios $C_{\rm sg}/C_{\rm bg}$  between the active split-gate and the back-gate electrodes which are respectively 0.83 for QPC$_1$ and 0.86 for QPC$_{2}$ and QPC$_{3}$. They are important quantities for the analysis of the QPC properties in the QH regime.
 
\newpage
\clearpage
\section{Fan diagram of bulk Landau levels}
\begin{figure*}[h!]
  	\centering
  	\includegraphics[width=\textwidth]{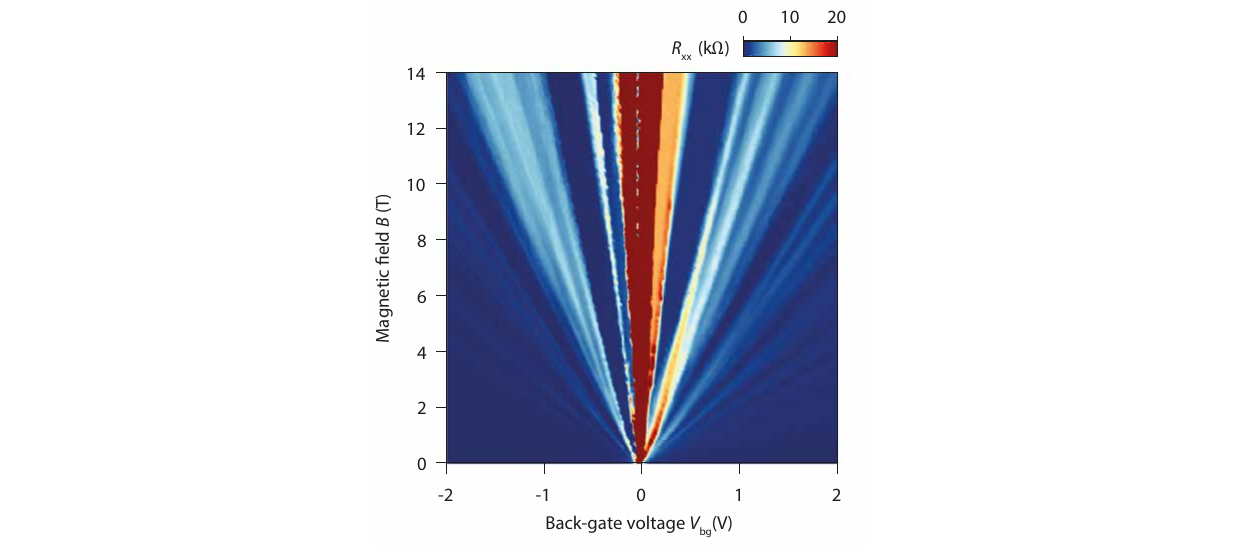}
  	\caption{\textbf{Landau fan diagram.} Longitudinal resistance $R_{\rm xx}$ of sample BNGr74 (device of the main text) versus back-gate voltage $V_{\text{bg}}$ and magnetic field $B$, measured at $0.02$~K.}
  	\label{Fan}
  \end{figure*}

In this section, we present the Landau fan diagram of sample BNGr74. Fig.~\ref{Fan} displays the longitudinal resistance $R_{\text{xx}}$ as a function of magnetic field $B$ and back-gate voltage $V_{\text{bg}}$, measured at $0.02$~K. This measurement was performed with a voltage $V_{\rm QPC}=+0.3$~V applied on each QPC to compensate the hole doping induced by the palladium split gates and ensure a quasi-homogeneous charge carrier density in the graphene layer. 
 
Broken-symmetry states in electron(hole)-type Landau levels emerge as minima in $R_{\text{xx}}$ above 5~T (3~T), consistent with the mobility $\mu = 130\, 000 \rm \, cm^2.V^{-1}.s^{-1}$ obtained for a charge carrier density of $1 \times 10^{12}\, \rm cm^{-2}$ from Hall measurements. In addition, an insulating behaviour develops  at charge neutrality with increasing magnetic field. The full-lifting of the degeneracies in the zeroth Landau level occurs above 4 T, allowing to perform interferometry experiments with the inner or outer electron edge channels of the zeroth Landau level at relatively low magnetic field values (see section \ref{sec:LowField}). 

From the position of the $R_{\text{xx}}$ minima, we extract a back-gate capacitance $C_{\rm bg}=1.45~\rm mF/m^2$ consistent with the bottom hBN thickness and a hBN dielectric constant $\epsilon_{\rm r}^{\rm BN}\approx 3$.

	\newpage
\clearpage
\section{Characterization of the QPCs in the quantum Hall regime}\label{section_QPC_14T}
QH interferometry experiments require a precise knowledge of the edge-channels configuration in the bulk of graphene, beneath the split-gate electrodes and in the split-gate constrictions. This section describes the action of the split-gate electrodes in the QH regime, which allows to determine the gate-voltage set points for the (partial) QPC pinch-off and tuning of QH edge channel transmissions. 

\begin{figure}[h!]
	\centering
	\includegraphics[width=1\textwidth]{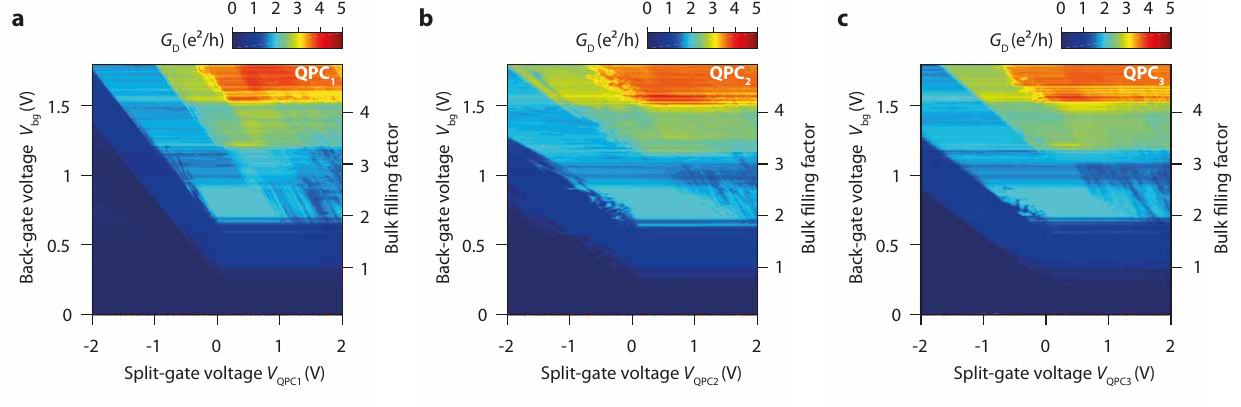}
	\caption{\textbf{QPC conductance maps at $\boldsymbol{14}$ T.} \textbf{a, b, c,} Diagonal conductance $G_{\text{D}}$ versus split-gate voltages, $V_{\text{QPC}}$, and back-gate voltages, $V_{\text{bg}}$, for the three QPCs of the device presented in the main text. During a measurement, only one QPC is studied and the two other sets of split gates are kept floating. The slope of the diagonal stripes corresponds to the capacitance ratio between the QPC constriction and the back gate. This slope is about twice/three times smaller than the zero-field slope for QPC$_2$ and QPC$_3$, but is only slightly smaller for QPC$_1$ (due to the unintentional absence of gap between the two electrodes of this QPC).}
	\label{fig:QPC_14T_Gd}
\end{figure}

\begin{figure}[h!]
	\centering
	\includegraphics[width=1\textwidth]{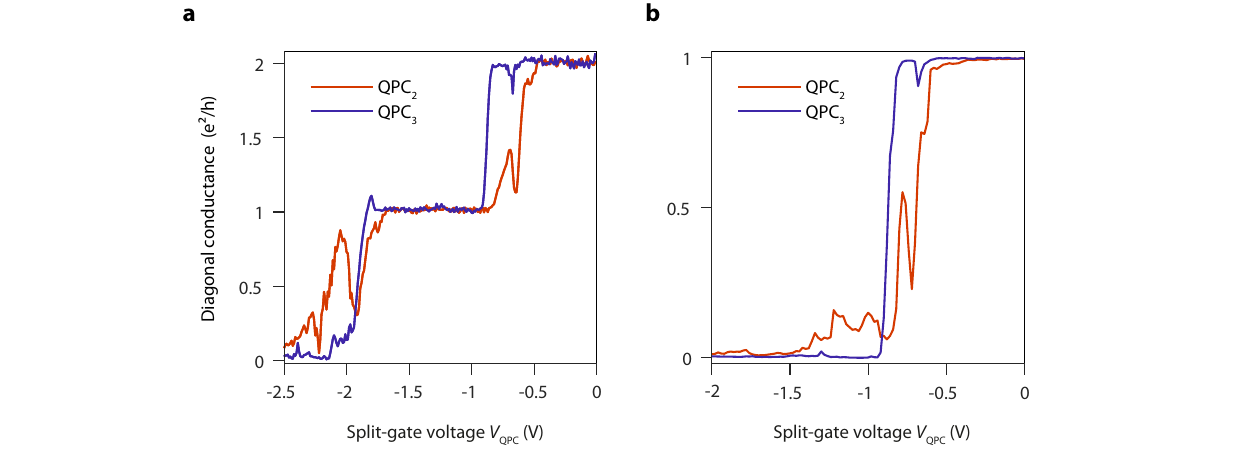}
	\caption{\textbf{QPC transmission curves at $\boldsymbol{14}~\text{T}$.}  Evolution of the diagonal conductance $G_{\text{D}}$ as a function of split-gate voltages $V_{\text{QPC}}$ at fixed back-gate voltage $V_{\text{bg}}$. \textbf{a,} $V_{\text{bg}} =0.88$ V. \textbf{b,}  $V_{\text{bg}} = 0.53$ V.} 
	\label{fig:QPC_14T_linecut}
\end{figure}

Figure~\ref{fig:QPC_14T_Gd} displays the diagonal conductance $G_{\text{D}}$ as a function of split-gate and back-gate voltages, $V_{\text{QPC}}$ and $V_{\text{bg}}$, for the three QPCs. The three conductance maps show  features similar to those reported in ref.~\cite{Zimmermann2017} for a QPC operating in the QH regime. At negative split-gate voltages, $G_{\text{D}}$ draws diagonal strips of nearly constant and quantized values. They have a smaller slope than the zero-field diagonal lines of Fig.~\ref{fig:QPC_0T}, indicative of the smaller couplings at the constrictions characterized by capacitance ratios $C_{\rm QPC}/C_{\rm bg} \simeq 0.58$, 0.31 and 0.36 for QPC$_{1}$, QPC$_{2}$ and QPC$_{3}$, respectively.  As demonstrated in ref.~\cite{Zimmermann2017}, the quantized $G_{\text{D}}$ values indicate the number of transmitted QH edge channels through the QPC. For a given bulk filling factor, the QH edge channels can be backscattered by applying a negative split-gate voltage $V_{\text{QPC}}$. For instance, at $V_{\text{bg}}=0.75$~V, the bulk filling factor is $\nu \simeq 2$, leading to the blue rhombi of $G_{\text{D}}=2e^2/h$ located near $V_{\text{QPC}}=0$~V in Fig.~\ref{fig:QPC_14T_Gd}. Decreasing $V_{\text{QPC}}$ to negative values, the conductance drops to the dark blue strip of $G_{\text{D}}=e^2/h$, and then to $G_{\text{D}}=0$ at even more negative values. These conductance changes reflect the successive backscattering of the QH edge channels at the QPC~\cite{Zimmermann2017}.
The linecuts of Fig.~\ref{fig:QPC_14T_linecut} further illustrate such a successive pinch off of the inner and outer edge channels at $\nu \simeq 2.5$ (Fig.~\ref{fig:QPC_14T_linecut}a) and the pinch off the outer edge channel at $\nu \simeq 1.5$ (Fig.~\ref{fig:QPC_14T_linecut}b).

As discussed in section \ref{section_QPC_0T}, the hole-doped graphene regions beneath  the non active split-gate electrodes intervene in the transmission of the whole device when studying the properties of a particular split gate. These hole-doped regions have a lower filling factor than the bulk and can therefore backscatter some bulk QH edge channels. As a consequence, the QH plateaus as a function of back-gate voltage at $V_{\text{QPC}}\sim 0$~V in the QPC maps of Fig.~\ref{fig:QPC_14T_Gd} are not centered at the integer bulk filling factors indicated on the right axis  and determined by the fan diagram $R_{\rm xx}(V_{\rm bg},B)$ at compensated split-gate voltages (see Fig.~\ref{Fan}). 

\begin{figure}[h!]
	\centering
	\includegraphics[width=1\textwidth]{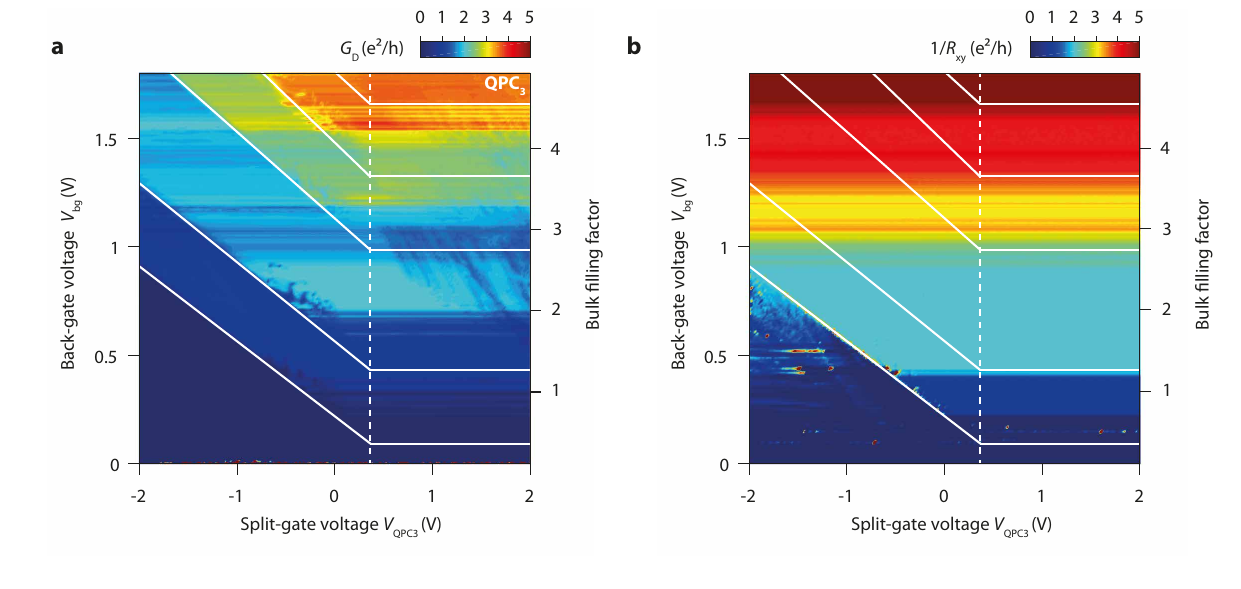}
	\caption{ \textbf{QPC map compared to Hall resistance map at $\boldsymbol{14}$ T.} \textbf{a,} Diagonal conductance $G_{\text{D}}$ versus split-gate voltage, $V_{\text{QPC}}$, and back-gate voltage, $V_{\text{bg}}$, for QPC$_3$. \textbf{b,} Inverse of the transverse Hall resistance $1/R_{\text{xy}}$ versus $V_{\text{QPC}}$ and $V_{\text{bg}}$. The vertical dashed white line indicates the split-gate voltage that compensates the hole doping induced by the split-gate electrodes (iso-density in the bulk and beneath the active split gate). This voltage is determined in the zero-field maps of Fig.~\ref{fig:QPC_0T} at the intersection between the diagonal line and the main horizontal line of the bulk charge neutrality point. The horizontal solid white lines delineate the quantized plateaus in the Hall resistance that are centered at integer bulk filling factors (indicated on the right axis). The diagonal lines delineate the diagonal strips of constant $G_{\text{D}}$ in the QPC map, that is, conductance plateaus given by the number of transmitted edge channels through the QPC (see ref.~\cite{Zimmermann2017} for a detailed analysis). For consistency, these diagonal lines meet the horizontal ones of the bulk Hall resistance right at their intersect with the vertical line.}
	\label{fig:QPC_Hall_lines_v2}
\end{figure}

The comparison in Fig.~\ref{fig:QPC_Hall_lines_v2} between a QPC map and the transverse Hall resistance $R_{\text{xy}} $ that relates to the bulk filling factor bears out this observation. The $\nu=2$ plateau develops at lower back-gate voltage in the Hall resistance than in the diagonal resistance across the QPCs. Despite the fact that the bulk has two QH edge channels when $1/R_{\text{xy}} = \frac{2e^2}{h}$ at, for instance,  $V_{\text{bg}} = 0.5$ V,  the non active QPCs that have lower filling factors backscatter the inner edge channel leading to $G_{\text{D}} =  e^2/h$ in the QPC conductance map. 

Furthermore, for the data presented in the main text, we assessed the number of bulk QH edge channels through the value of the Hall resistance plateau. For all figures of the main text, we measured  $1/R_{\text{xy}} = \frac{2e^2}{h}$, which indicates that two edge channels propagate in the graphene bulk.

\newpage
\clearpage

   \section{Aharonov-Bohm oscillations at positive plunger-gate voltage}
	\label{Section5}
\begin{figure}[h!]
   	\centering
   	\includegraphics[width=1\textwidth]{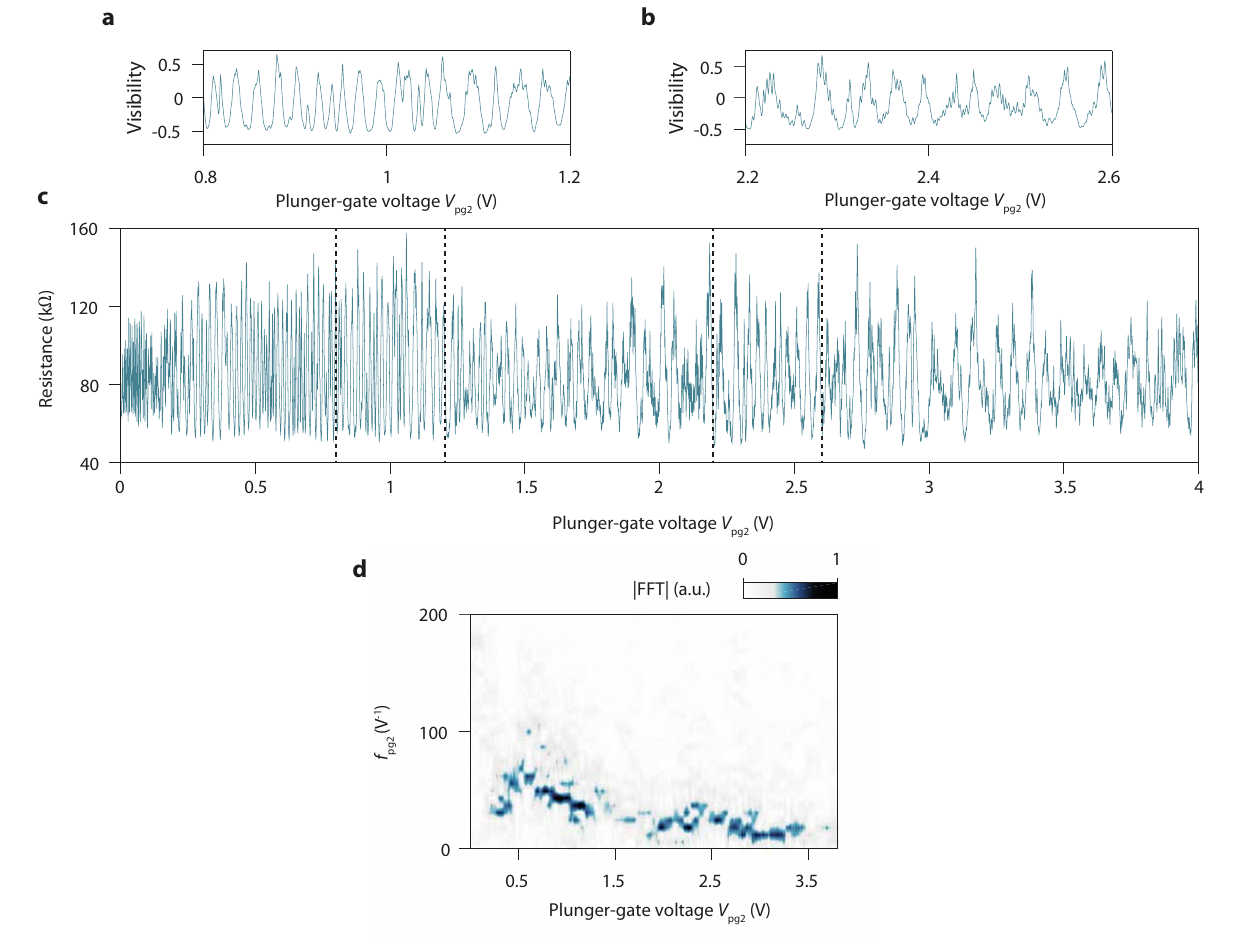}
   	\caption{\textbf{Resistance oscillations at positive plunger-gate voltage} \textbf{a, b, c,} Resistance oscillations as a function of plunger-gate voltage $V_{\text{pg2}}$ measured in the small interferometer for $V_{\text{pg2}}>0$. a and b show zooms on smaller $V_{\text{pg2}}$ ranges of the resistance oscillations converted in visibility $(R-\bar{R})/\bar{R}$, where  $\bar R$ is the resistance background. \textbf{d,} Fourier amplitude of the resistance oscillations in c as a function of $V_{\rm pg2}$ and the plunger-gate voltage frequency $f_{\rm pg2}$.}
   	\label{fig:Positive_osc_plunger_gate}	
   \end{figure}

   In Fig.~\ref{fig:Positive_osc_plunger_gate}, we show the extension of the measurements performed in Fig.~2c of the main text to positive plunger-gate voltage. This corresponds to the accumulation of localized electron states beneath the plunger gate (see inset in Fig.~2c). Resistance oscillations appear clearly on the positive voltage range but with an irregular shape contrary to the oscillations at $V_{\text{pg2}}<0$~V. The Fourier amplitude of these resistance oscillations is shown in  Fig.~\ref{fig:Positive_osc_plunger_gate}. The frequency of the oscillations evolves non-monotonously and is poorly defined for some voltage ranges. 

\newpage
\clearpage
 \section{Aharonov-Bohm oscillations for different configurations of magnetic field and edge channels}
 
\begin{figure}[h!]
 	\centering
 	\includegraphics[width=\textwidth]{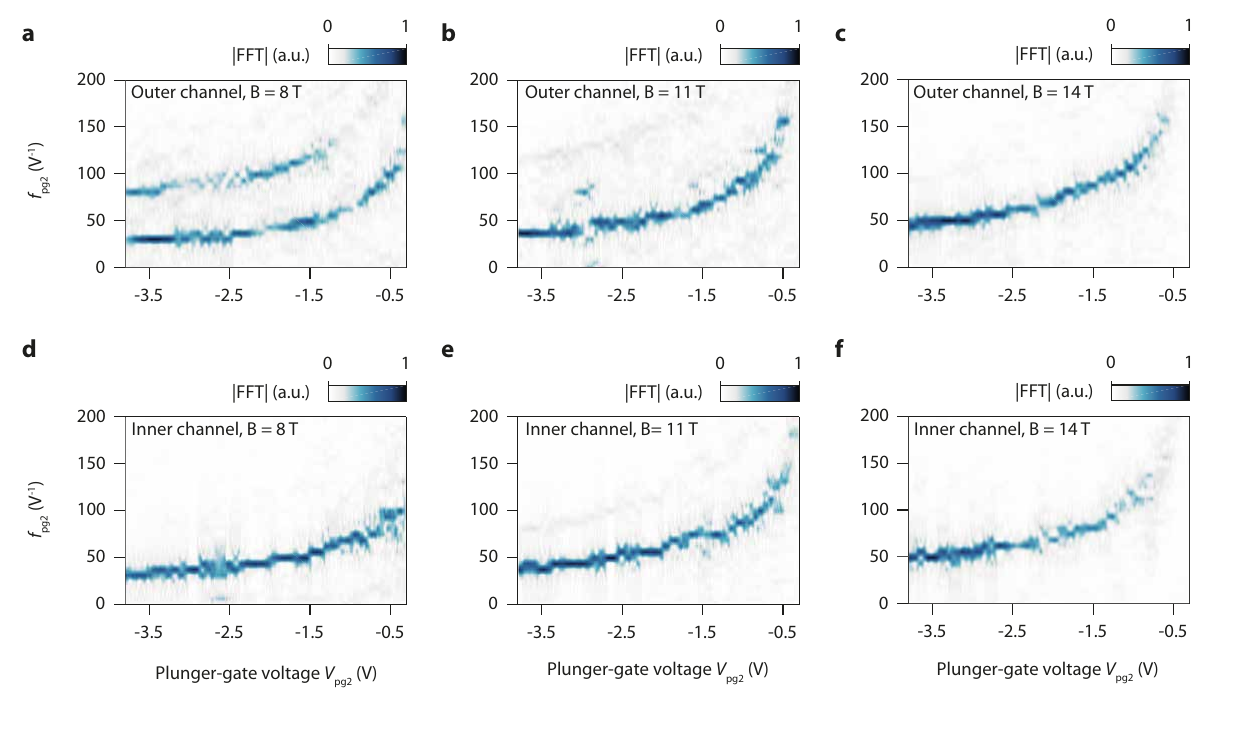}
 	\caption{\textbf{Fourier amplitude of the resistance oscillations.}  Fourier amplitude of the resistance oscillations observed in the small interferometer for different configurations of magnetic field and interfering edge channel, as a function of plunger-gate voltage $V_{\text{pg2}}$ and frequency $f_{\text{pg2}}$.}
 	\label{fig:FFT_various_config}
 \end{figure}

 In this section we present plots of the Fourier amplitude of the resistance oscillations with $V_{\text{pg2}}$ for experiments performed with different interfering edge channels and magnetic fields. In every cases, the frequency of the oscillations $f_{\text{pg2}}$ is well defined and shows a clear and continuous decrease while lowering $V_{\text{pg2}}$. As expected for the Aharonov-Bohm regime, the frequency of the oscillations increases with the magnetic field at fixed plunger-gate voltage whereas it does not change with the interfering edge channel. A significant component oscillating at twice the Aharonov-Bohm frequency is visible on Fig. \ref{fig:FFT_various_config}a. In this case, only the lowest frequency component was used to plot Fig. 2e in the main text.

\newpage
\clearpage
\section{Electrostatics of the plunger gate}

\begin{figure}
\begin{center}
\includegraphics[width=14cm,trim={0 0 0 0},clip]{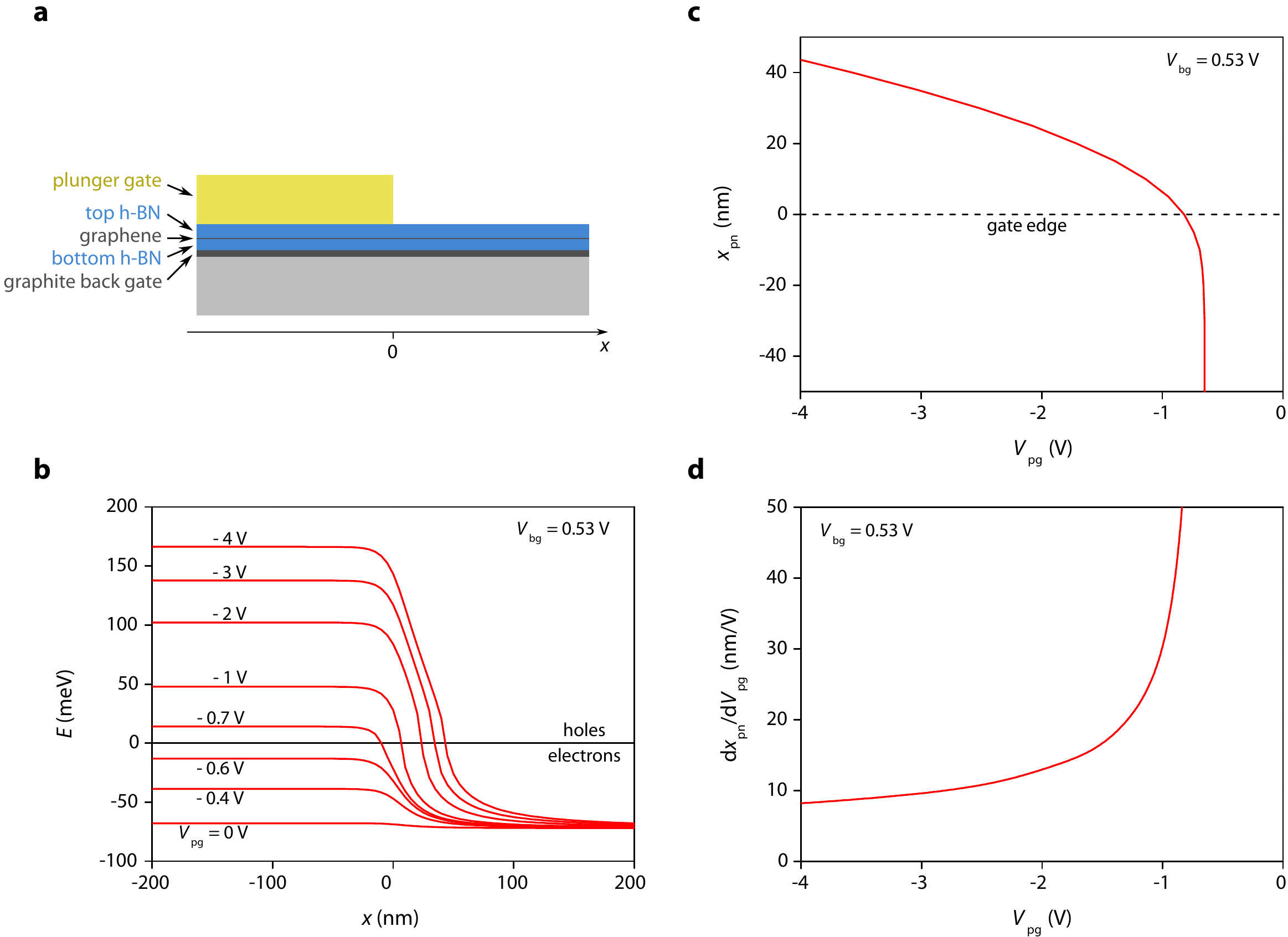}
\caption{\textbf{Plunger-gate electrostatics.} \textbf{a,} Schematics of the hBN/graphene/hBN heterostructure deposited on the graphite back gate and partially covered by the metallic plunger gate used to tune the interfering path length. \textbf{b,} Self-consistent electrostatic energy profiles $E=-eV$ in the graphene layer for a back-gate voltage $V_{\rm bg}=0.53$~V and plunger-gate voltages $V_{\rm pg}$ varying from 0 to $-4$~V. \textbf{c,} Position of the pn interface with respect to the gate edge as a function of the plunger-gate voltage. \textbf{d,} Displacement rate of the pn interface calculated as its derivative with respect to the plunger-gate voltage.} 
\label{fig:Plunger_electrostatics}
\end{center}
\end{figure}

The potential profile in the graphene below the plunger gate is determined by self-consistent electrostatic simulations in the vertical 2D plane shown in Fig.~\ref{fig:Plunger_electrostatics}a assuming translational invariance in the third direction. The simulation is done for the same hBN thicknesses as in the device of the main text, with $d_{\rm bottom}=18$~nm for the bottom layer and $d_{\rm top}=22$~nm for the top layer. The hBN dielectric permittivity $\epsilon_{\rm r}^{\rm BN}\approx3$  is extracted from the position of quantum Hall plateaus versus back-gate voltage.
The graphite back-gate is treated as a perfect metal. The graphene sheet is modelled by a charge density $\sigma(x)$ linked to the electrostatic potential $V(x)$ by the relation:
$$ \sigma(x) = (-e)\,{\rm sign}\big(V(x)\big)\,\frac{e^2 V(x)^2}{\pi\hbar^2v_{\rm F}^2} $$
where $v_{\rm F}=10^6$~m/s is the Fermi velocity in graphene. The electrostatic problem is solved self-consistently using a modified version of MaxFEM (http://www.usc.es/en/proxectos/maxfem), an electromagnetic simulation software based on the finite-element method. The mesh grid computed using Gmsh (http://gmsh.info) extends 1~$\mu$m in vertical and $2~\mu$m in horizontal.

The self-consistent solution $V(x)$ can be calculated for a given back-gate voltage $V_{\rm bg}$ and a series of plunger-gate voltages $V_{\rm pg}$ in order to determine the dependence of the pn interface position $x_{\rm pn}$ on the plunger-gate voltage. Equivalently, the local plunger-gate capacitance $C_{\rm pg}(x)$ can be extracted from a single self-consistent simulation (for example at $V_{\rm bg}=0$ and $V_{\rm pg}=-1$~V) using the quantum capacitance model \cite{Liu2013}. This model is based on the relation between $\sigma(x)$ and $V(x)$ given above, together with the definition of the capacitive couplings:
$$ \sigma(x)=-C_{\rm bg}\,\big(V_{\rm bg}-V(x)\big)-C_{\rm pg}(x)\,\big(V_{\rm pg}-V(x)\big) $$
where $C_{\rm bg}=\epsilon_0\epsilon_{\rm r}^{\rm BN}/d_{\rm bottom}$. This approach based on the determination of the local capacitance $C_{\rm pg}(x)$ has the advantage to provide the self-consistent solution for any set of back-gate and plunger-gate voltages without the need to solve again the full electrostatic problem.

The spatial variation of the potential energy $E(x)=-eV(x)$ below the plunger gate is plotted in Fig.~\ref{fig:Plunger_electrostatics}b for a fixed back-gate voltage $V_{\rm bg}=0.53$~V and various negative plunger-gate voltages corresponding to the experiment reported in Fig.~2 of the main text. The position $x_{\rm pn}$ of the pn interface with respect to the gate edge is plotted in Fig.~\ref{fig:Plunger_electrostatics}c as a function of the plunger-gate voltage, showing the following behavior: the formation of the pn interface occurs at $V_{\rm pg}=-0.65$~V (in the data this happens around $\simeq-0.3$~V instead, due to the hole doping of $+0.38$~V from the palladium split-gate electrodes, corresponding to the charge neutrality point below the plunger gate), then the fast displacement of the pn interface corresponds to the expulsion of the pn interface from below the plunger gate, and finally the pn interface moves slower and slower for large negative plunger-gate voltages. The displacement rate $\frac{{\rm d}x_{\rm pn}}{{\rm d}V_{\rm pg}}$ plotted in Fig.~\ref{fig:Plunger_electrostatics}d is used in the main text to calculate the non-linear lever arm $\alpha=L_{\rm pg}\times\frac{{\rm d}x_{\rm pn}}{{\rm d}V_{\rm pg}}$ of the plunger gate with contour length $L_{\rm pg}$. This lever arm provides the theoretical conversion between plunger-gate voltage and interferometer area, which writes $\Delta A = \alpha \, \Delta V_{\text{pg}}$, and which is compared in Fig.~2f with the oscillation frequency measured experimentally. $L_{\rm pg}$ remains an adjustable parameter because the position of the graphene edges is known with an uncertainty of $\pm150$ nm. To reproduce the measurement, a plunger-gate contour $L_{\rm pg}=1.8 \ \mu$m is used, in good agreement with the expected lithographic length of $1.5\pm0.3\ \mu$m (the uncertainty of the graphene edge position contributes twice).

\newpage
\clearpage

  \section{Aharonov-Bohm oscillations in the medium interferometer}
  
\begin{figure}[h!]
 \centering
 \includegraphics[width=\textwidth]{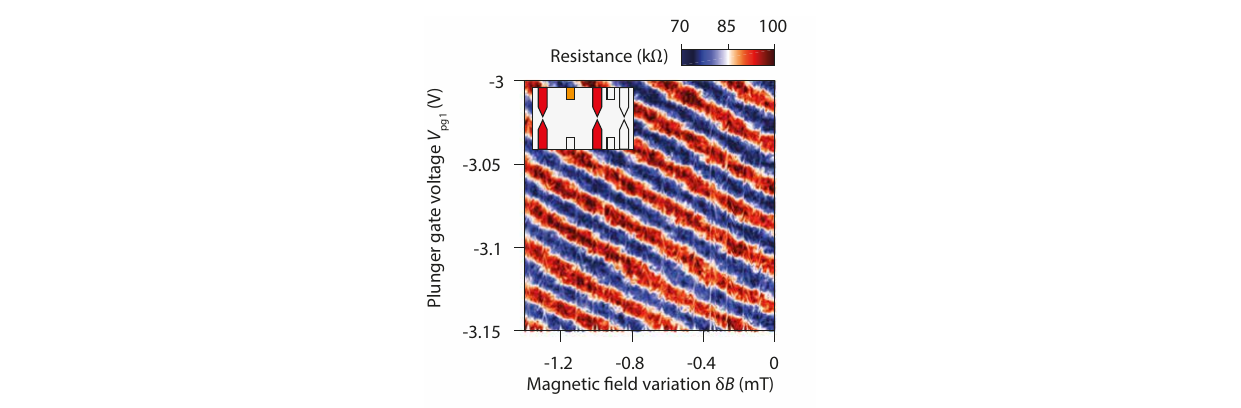}
\caption{\textbf{Aharonov-Bohm oscillations in the medium interferometer.} Diagonal resistance as a function of plunger-gate voltage $V_{\text{pg2}}$ and magnetic field variation $\delta B$ in the medium interferometer measured at 14 T with the outer edge channel interfering. The inset schematic indicates the active QPCs (in red) and plunger gate (in orange).}
\label{Pyjama_nu1_MI}
\end{figure}

To complement the $(\delta B,V_{\rm pg})$ maps shown in Fig. 3a and b for the small and large interferometers, we present in Fig.~\ref{Pyjama_nu1_MI} the map obtained for the medium interferometer, in the same conditions, i.e. with the outer edge state at $B=14$~T. The constant resistance lines have a negative slope indicating the Aharonov-Bohm origin of the oscillations. The field periodicity is 0.40 mT corresponding to an Aharonov-Bohm area of 10.4 $\mu \rm m^2$ in good agreement with the expected lithographic area (see Table \ref{Table2}).  
\newpage
 \section{Interferometry experiments with inner edge state at 14 T in the three interferometers}
 
\begin{figure}[h!]
 \centering
 \includegraphics[width=\textwidth]{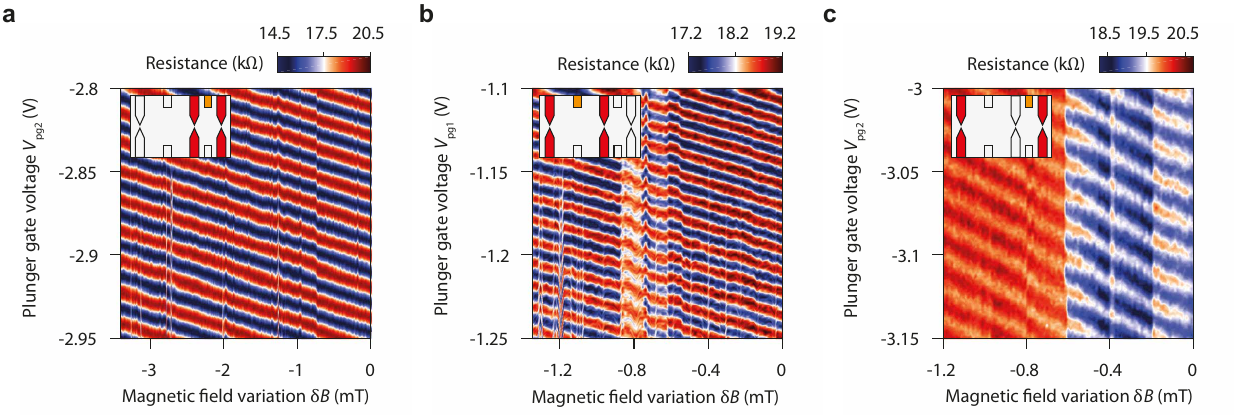}
\caption{\textbf{Aharonov-Bohm oscillations with the inner edge channel.} \textbf{a, b, c,} Diagonal resistance versus plunger-gate voltage $V_{\text{pg1,2}}$ and magnetic field $\delta B$ for the small, medium and large interferometers, respectively, with the inner edge channel interfering at 14 T. The inset schematics indicate the active QPCs (in red) and plunger gates (in orange).}
\label{fig:Pyjama_nu2}
\end{figure}

In this section we present additional interferometry experiments performed with the inner edge channel of the zeroth Landau level at $B=14$ T. Fig.~\ref{fig:Pyjama_nu2}a, b and c show the diagonal resistance of the device as a function of plunger-gate voltages and magnetic field for the small, medium and large interferometers, respectively.  The results are virtually identical to those performed with the outer edge channel. The magnetic field periods extracted from these measurements are respectively of 1.23, 0.39 and 0.27 mT.

\newpage
\clearpage

\section{Interferences at lower magnetic fields}\label{sec:LowField}

\begin{figure}[h!]
     	\centering
     	\includegraphics[width=0.8\textwidth]{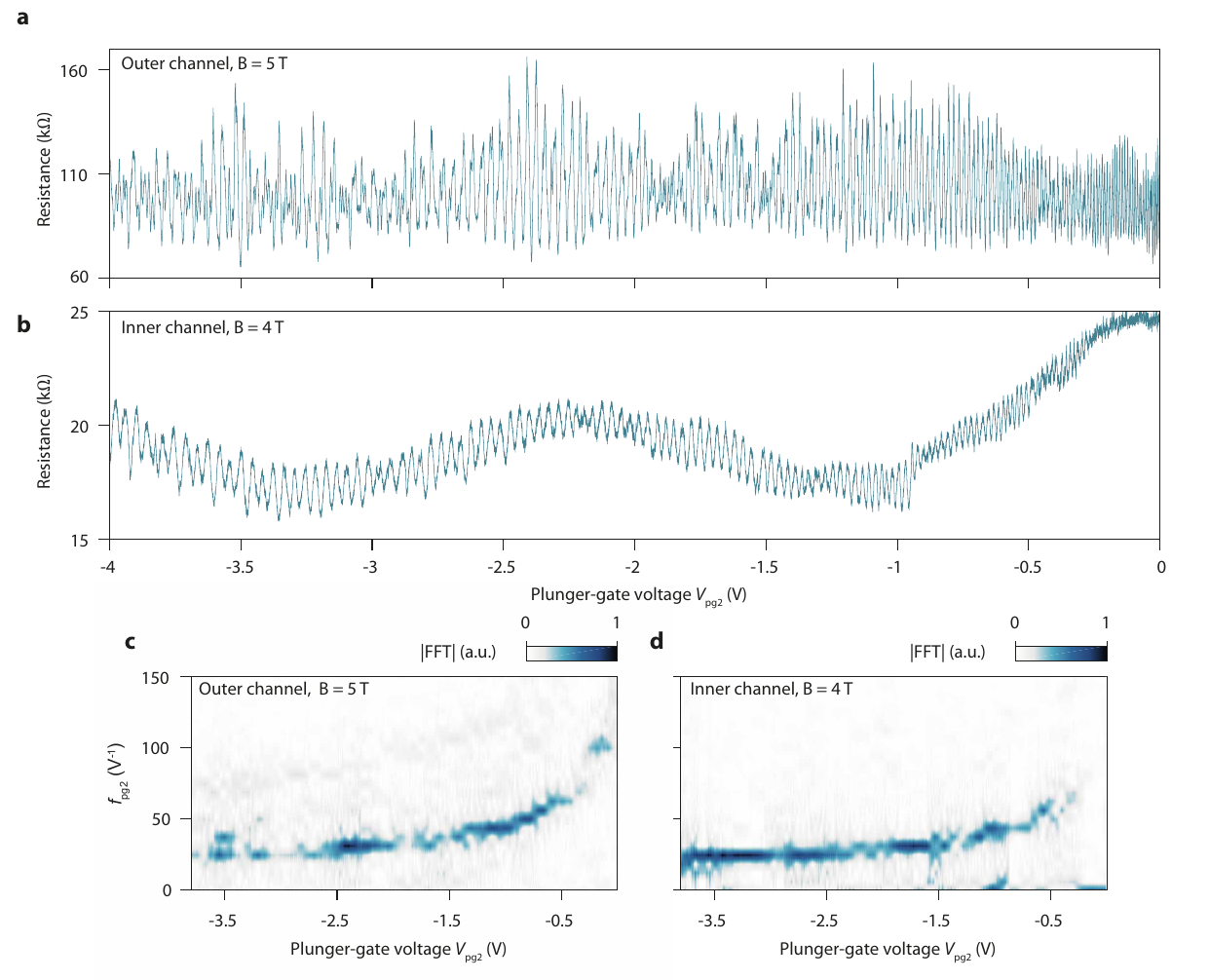}
     	\caption{\textbf{Resistance oscillations at low magnetic fields.} \textbf{a, b,} Resistance oscillations as a function of plunger-gate voltage $V_{\text{pg2}}$ measured in the small interferometer at $5$~T with the outer edge channel, and $4$~T with the inner edge channel, respectively. \textbf{c, d,} Fourier amplitude of the resistance oscillations in a and b.}
     \label{fig:Low_field}
     \end{figure}

Here we show that the device BNGr74 presented in the main text can also operate at low magnetic field. Stable Aharonov-Bohm interference were observed with the outer and inner edge channels respectively down to 5~T and 4~T as displayed in Fig.~\ref{fig:Low_field}a and b. The respective Fourier amplitudes of the resistance oscillations are shown in Fig.~\ref{fig:Low_field}c and d.  

\newpage

\section{Aharonov-Bohm oscillations in other devices}\label{OtherDevices}
	In this section we present the data obtained on two other devices, BNGr64 and BNGr30. They do not have a graphite back gate, and the silicon substrate serves as the back gate instead. Even without graphite electrode, we observed for both samples Aharonov-Bohm oscillations, indicating that the absence of charging effect is not only related to the screening by the graphite gate. 
	
  \subsection*{BNGr64 device}
	
  \label{BNGr64}
\begin{figure}[h!]
     	\centering
     	\includegraphics[width=0.75\textwidth]{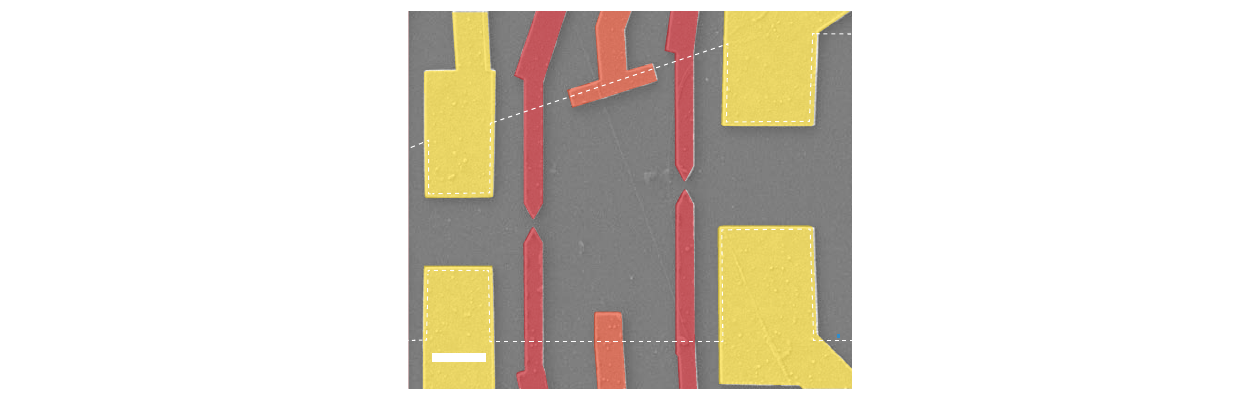}
     	\caption{\textbf{QH-FP interferometer in sample BNGr64.} False-colored scanning electron micrograph of the device. Graphene edges are represented by the white dotted line. Contacts, QPCs and plunger gates are color-coded in yellow, red and orange. Scale bar is 1 $\rm \mu m$.}
     \label{fig:BNGr64_SEM}
     \end{figure}

We first present the data for the device BNGr64 shown in Fig.~\ref{fig:Image_samples}b. In this device,  three out of four QPCs were operating correctly enabling us to perform experiments with only one of the two interferometers, whose scanning electron micrograph is displayed in Fig.~\ref{fig:BNGr64_SEM}. This device was studied using a larger ac bias-voltage excitation of 20 $\mu$V and using the bottom plunger gate. The large plunger gate was kept grounded during the measurements.

\begin{figure}[h!]
     	\centering
     	\includegraphics[width=0.85\textwidth]{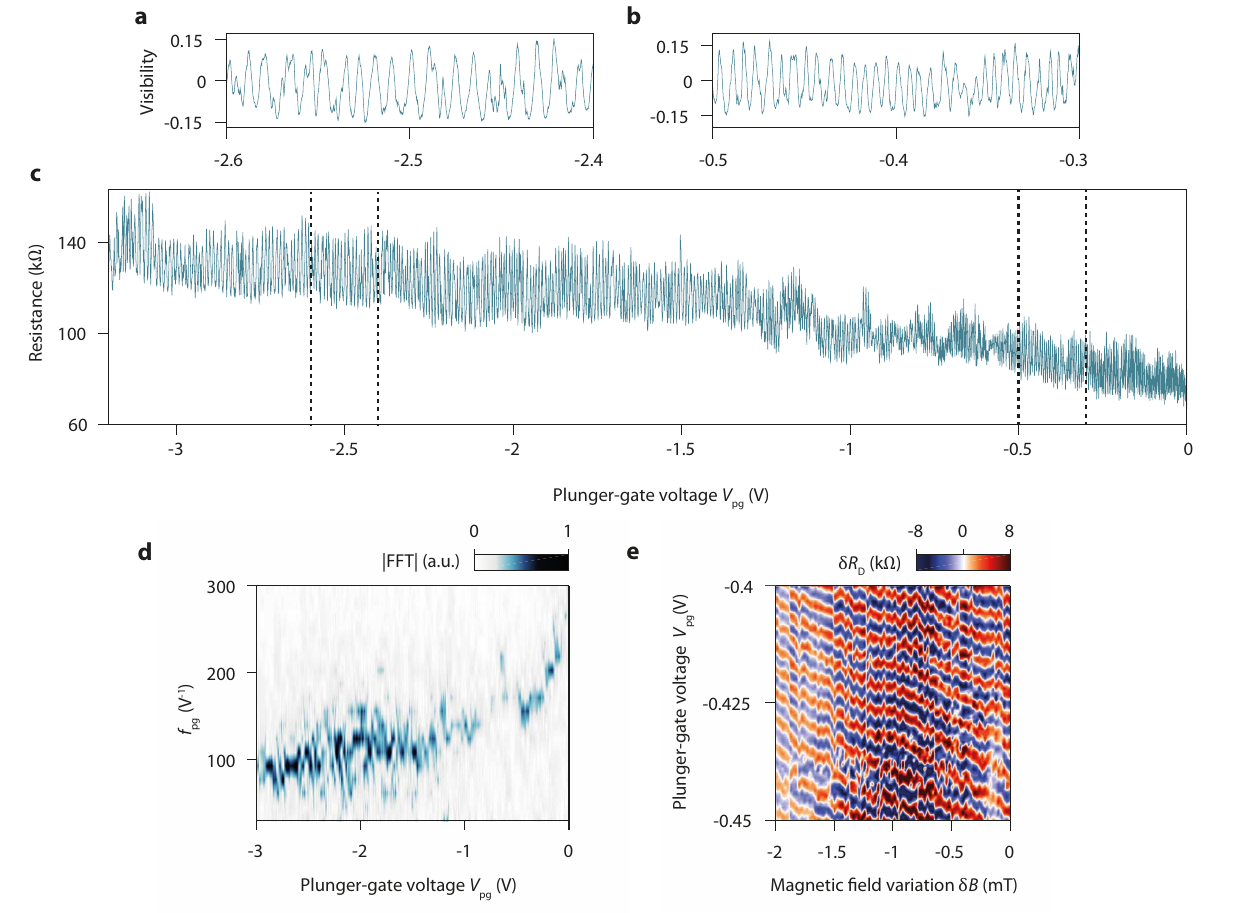}
     	\caption{\textbf{Resistance oscillations in sample BNGr64.} \textbf{a, b, c,} Resistance oscillations  induced by change of the plunger-gate voltage $V_{\rm pg}$  in interferometry experiments with the inner edge channel at 14 T. Clear resistance oscillations are visible lowering $V_{\rm pg}$, on top of a continuous increase of the mean resistance of the device evidenced in c. a and b show zooms on smaller $V_{\text{pg}}$ ranges of the resistance oscillations converted in visibility $(R-\bar{R})/\bar{R}$, where  $\bar R$ is the resistance background. \textbf{d,} Amplitude of the Fourier transform of resistance oscillations presented in c with respect to the plunger-gate voltage $V_{\text{pg}}$ and the frequency $f_{\text{pg}}$. A continuous decrease of the oscillations frequency is observed while decreasing $V_{\text{pg}}$. \textbf{e,} Evolution of the resistance oscillations  as function of the plunger-gate voltage $V_{\text{pg}}$  and the magnetic field variation $\delta B$ after subtraction of a resistance background for each plunger-gate voltage sweep. Constant $\delta R_{\rm D}$ lines have a negative slope characteristic of oscillations induced by Aharonov-Bohm effect.}
     \label{fig:BNGr64_data}
     \end{figure}

We present interferometry experiments performed with the outer interfering edge channel at 14 T with a bulk filling factor $\nu_{b}=1.1$. Contrary to the data presented in the main text, there is only one electron-like edge channel propagating in the interferometer. Fig.~\ref{fig:BNGr64_data}c shows the evolution of the diagonal resistance with plunger-gate voltage $V_{\rm pg}$. Clear resistance oscillations are observed while decreasing $V_{\rm pg}$ from 0 to -3.2 V. Contrary to the data presented in Fig.~2c of the main text, the oscillations show many phase shifts as well as some visibility losses, reflecting the lower degree of stability and coherence of the device. The visibility of the oscillations is typically about 15\% as evidenced in Fig.~\ref{fig:BNGr64_data}a and b. The Fourier transform amplitude of the oscillations is presented in Fig.~\ref{fig:BNGr64_data}d and shows a decrease of the frequency of the oscillations $f_{\text{pg}}$ with the plunger-gate voltage $V_{\text{pg}}$ consistent with that in Fig.~2 of the main text.
 
The evolution of the diagonal resistance oscillations with both the plunger-gate voltage and the magnetic field in this configuration is shown in Fig.~\ref{fig:BNGr64_data}e. A smooth resistance background for each sweep was subtracted to evidence lines of constant Aharonov-Bohm phase and get rid of average-conductance variations. Constant resistance values form lines with a negative slope in the $\delta B$--$V_{\text{pg}}$ plane which shows that this device operates in the Aharonov-Bohm regime. From these measurements, we extract a magnetic field period of $0.42$ mT corresponding to an enclosed area of $9.9 \ \mu\text{m}^2$ in agreement with the geometrical surface of 11.5 $\mu\text{m}^2$.

   \newpage
 \subsection*{BNGr30 device}
\label{BNGr30}   
 
\begin{figure}[h!]
     	\centering
     	\includegraphics[width=0.75\textwidth]{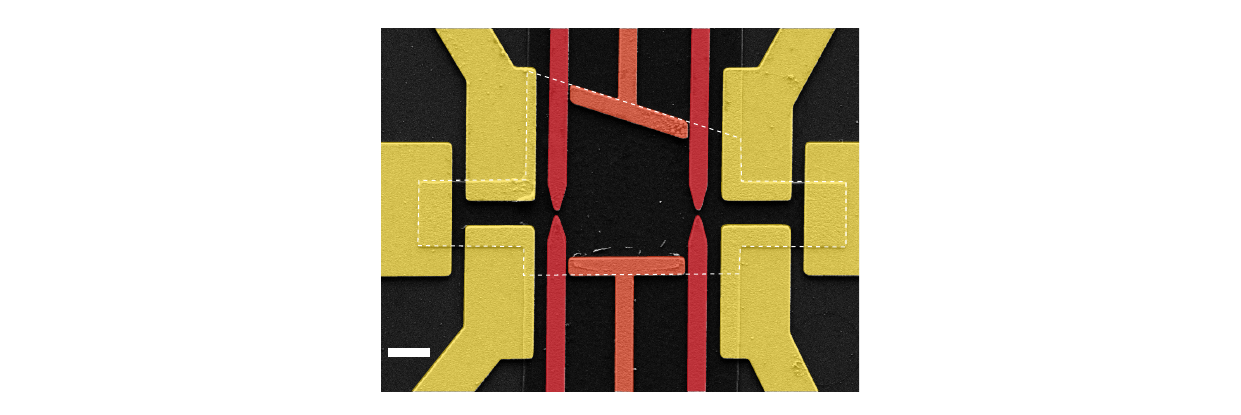}
     	\caption{ \textbf{QH-FP interferometer in sample BNGr30.} False-colored scanning electron micrograph of the device. Graphene edges are represented by the white dotted line. Contacts, QPCs and plunger gates are color-coded in yellow, red and orange, respectively. Scale bar is 1 $\rm \mu m$.}
     \label{fig:BNGr30_SEM}
     \end{figure}

Here we present the data for the device BNGr30, displayed in Fig.~\ref{fig:Image_samples}c. Contrary to the two previous samples, before the deposition of the metallic contacts and of the gates, the heterostructure was etched and shaped using a hard-mask of HSQ resist to uncover the graphene edges at determined positions. After a second e-beam lithography steps, both the contacts and the split gates were made by depositing a Cr/Au bilayer. In this device, the plunger gates cover nearly all the graphene edges between the two QPCs. A scanning electron micrograph of the device is shown in Fig.~\ref{fig:BNGr30_SEM}.
   
\begin{figure}[h!]
     	\centering
     	\includegraphics[width=0.85\textwidth]{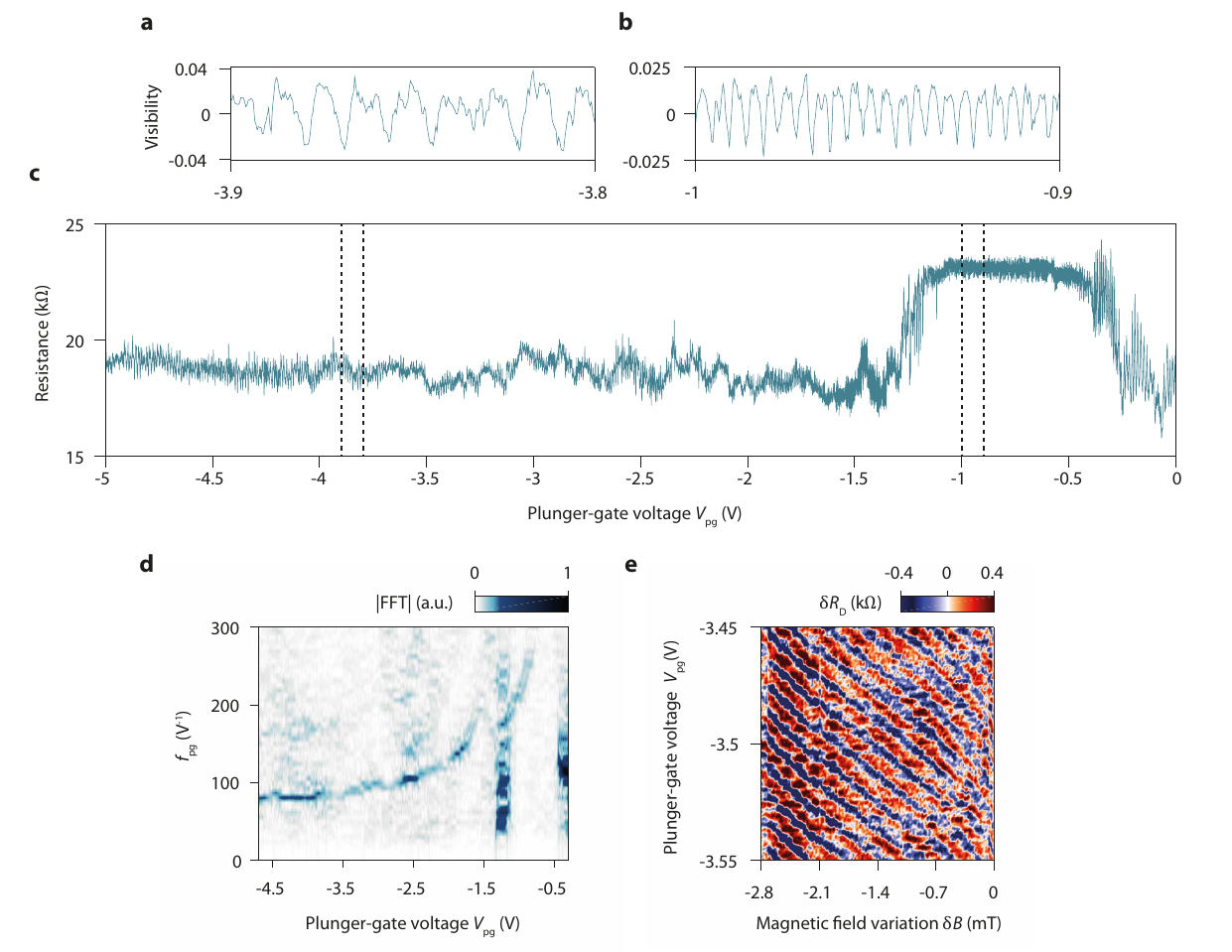}
     	\caption{\textbf{Resistance oscillations in sample BNGr30.} \textbf{a, b, c,} Resistance oscillations  induced by a change of the plunger-gate voltage $V_{\rm pg}$  in interferometry experiments with the inner edge state at 14 T. The abrupt change in c of the mean resistance value at $V_{\rm pg}\approx-1.2$ V and $V_{\rm pg}\approx-0.2$ V might originate from instability of the QPCs. a and b show zooms on smaller $V_{\text{pg}}$ ranges of the resistance oscillations converted in visibility $(R-\bar{R})/\bar{R}$, where  $\bar R$ is the resistance background. \textbf{d,} Amplitude of the Fourier transform of resistance oscillations presented in c with respect to the plunger-gate voltage $V_{\text{pg}}$ and the corresponding voltage frequency $f_{\text{pg}}$. A continuous decrease of the oscillations frequency is observed while decreasing $V_{\text{pg}}$.  The divergence at $V_{\rm pg}\approx-1.2$ V is an artefact arising from the rapid change of the mean resistance value at this plunger-gate voltage. \textbf{e,} Evolution of the resistance oscillations with both the plunger-gate voltage $V_{\text{pg}}$  and the magnetic field variation $\delta B$ after subtraction of a resistance background for each plunger-gate voltage sweep. Constant $\delta R_{\rm D}$ lines have a negative slope characteristic of oscillations induced by the Aharonov-Bohm effect.}
     \label{fig:BNGr30_data}
     \end{figure}

 Interferometry experiments performed in this device with the inner edge channel at bulk filling factor $\nu_{b} =2.3$ and 14 T are presented in Fig.~\ref{fig:BNGr30_data}. Resistance oscillations induced by a change of the top plunger-gate voltage $V_{\rm pg}$ are shown in Fig.~\ref{fig:BNGr30_data}a-c. They appear on the entire range of $V_{\text{pg}}$ voltage even though the stability of the QPC is affected by the value of $V_{\text{pg}}$. These oscillations have a small visibility typically varying between 2 and 5 \% as shown in Fig. \ref{fig:BNGr30_data}a and b. The Fourier transform analysis of the oscillations, shown in Fig.~\ref{fig:BNGr30_data}d reveals a similar lowering of the frequency $f_{\text{pg}}$ of the oscillations with the plunger-gate voltage (the absence of well-defined frequency for the oscillations at $V_{\text{pg}} \simeq -1.2$ V arises from the rapid drop of the resistance background).
 
  In Fig.~\ref{fig:BNGr30_data}e, we show the evolution of resistance oscillations with both the magnetic field and the plunger-gate voltage. The constant phase lines have a negative slope evidencing that the oscillations result from the Aharonov-Bohm effect. We can extract a magnetic field period of 0.37 mT corresponding to an area enclosed by the interfering edge state of 11.2 $\mu\rm m^2$ in good agreement with geometric area of 10.1 $\mu\rm m^2$.
 
 \newpage
 \clearpage


\section{Absence of charging effect}

Here we discuss the absence of Coulomb blockade in graphene FP interferometers. We follow the approach proposed in ref.~\cite{Halperin2011,Sivan2016} and estimate the relevant capacitances describing the electrostatics of the system. We use the notations of ref.\cite{Sivan2016}, make approximate calculations for the small interferometer with a graphite back gate and discuss the case of the devices without graphite back gate. These calculations allow us to evaluate the parameter  $\xi = \frac{C_{\rm eb}}{C_{\rm b}+C_{\rm eb}}$, where  $C_{\rm b}$ is the bulk-to-gate capacitance and $C_{\rm eb}$ the edge-to-bulk capacitance, which defines according to ref.~\cite{Halperin2011} if the device is operating in the Aharonov-Bohm or Coulomb-dominated regime.

\subsection*{Bulk capacitance $C_{\rm b}$}

The bulk capacitance $C_{\rm b}$ refers to the capacitance of the electrons located in the central part of the cavity and spatially separated from the conducting edge channels. These bulk electrons belong to the last partially-occupied Landau level and form an isolated island capacitively coupled to the gate electrodes~\cite{Rosenow07} (back gate, plunger gates, and split-gates) . The electrostatic coupling of the bulk to the interfering edge channel is considered separately in another capacitance term $C_{\rm eb}$ discussed later.

For our device with a graphite back gate, the bulk capacitance is mostly given by $C_{\rm b}=C_{\rm bg}A_{\rm geo}$ where $C_{\rm bg}=1.45~\text{mF/m}^2$ is the effective back-gate capacitance and $A_{\rm geo}$ is the geometrical area. For our small FP cavity, we  obtain $C_{\rm b}=4.5\times 10^{-15}$~F. The corresponding bulk charging energy is thus $E_{\rm C}=\frac{e^2}{2C_{\rm b}}=18~\mu$eV comparable to that reported for devices in GaAs heterostructures~\cite{Nakamura2019} in which AB oscillations with fractional edge channels were reported.

For our devices without graphite back gate, the 285 nm thick SiO$_2$ layer gives $C_{\rm bg}=0.12~\text{mF/m}^2$. Thus for devices having similar sizes, $C_{\rm b}$ is approximately decreased by a factor 10 with respect to devices with graphite back gate. For the two devices presented in section~\ref{OtherDevices},  BNGr64 and BNGr30, which respectively have geometrical surfaces of 11.5  and 10.1 $\mu\text{m}^2$, we obtain respectively $C_{\rm b}=1.4$ and $1.2\times10^{-15}$~F.

On top of this bulk-to-back-gate capacitance, one needs to add the contribution of the plunger-gate and split-gate electrodes resting atop the 20 nm thick capping hBN. This contribution is difficult to evaluate because the top gates are not located directly above the bulk island. However, they still provide an additional parallel capacitive coupling leading to an increase of $C_{\rm bg}$ and a reduction of the overall bulk charging energy. We note that this effect may play a significant role in devices on silicon substrate and may become the main contribution to the bulk capacitance.

\subsection*{Interfering edge channel capacitance $C_{\rm e}$}

Similarly, the interfering edge channel is capacitively coupled to gates electrodes and one can define a edge-to-gates capacitance $C_{\rm e}$. For a sake of completeness, we also evaluate it though it does not appear in the expression of $\xi$. $C_{\rm e}$ is the sum of two contributions : the edge-to-top-gates capacitance $C_{\rm e/tg}$ and the edge-to-back-gate capacitance $C_{\rm e/bg}$.

The latter can be evaluated following a similar approach as above. In this case, $C_{\rm e/bg} = 2 L w C_{\rm bg}$ where 2$L$ is the FP cavity perimeter and $w$ is the width of the compressible stripe corresponding to the QH edge channel. Assuming $w=l_{B}$ the magnetic length ($\approx 7$ nm at 14 T), we obtain $C_{\rm e/bg}=8.6\times10^{-17}~$F. This contribution is likely to be increased by edge-channel reconstruction\cite{Chklovskii1992}, which could occurs along the smooth potential of the pn-junctions.

On the other hand, $C_{\rm e/tg}=C_{\rm e/sg}+C_{\rm e/pg}$ is the sum of the capacitance $C_{\rm e/sg}$ between the split-gates and the interfering edge channel and the capacitance $C_{\rm e/pg}$ between the plunger gate and the interfering edge channel. The latter can be extracted from the plunger-gate voltage period $\Delta V_{\rm pg}$ of AB oscillations, as an oscillation corresponds to the addition/removal of one flux quantum inside the area enclosed by the edge channel and thus of an electron in the corresponding Landau level. Then, $C_{\rm e/pg}=\frac{e}{\Delta V_{\rm pg}}=1.6\times10^{-17}$~F for a typical voltage period $\Delta V_{\rm pg}=10$~mV.  Note that $C_{\rm e/pg}$ scales as the perimeter $L_{\rm pg}$ of the plunger gate (geometrically 1.5 $\mu$m). 
From this evaluation, we can also estimate $C_{\rm e/sg}$ by making the reasonable assumption that the electrostatics is the same for the split-gates and for the plunger gate. Thus $C_{\rm e/sg}=\frac{L_{\rm sg}}{L_{\rm pg}}C_{\rm e/pg}=4.7\times 10^{-17}~$F with $L_{\rm sg}=4.4~\mu$m is the total length of the split-gate electrodes defining the cavity. \
The total edge capacitance is thus about $C_{\rm e}=1.5\times10^{-16}~$F.


In devices with silicon back gate, we expect $C_{\rm e}$ to be lower due to a smaller  $C_{\rm bg}$, but still of the same order of magnitude.

\subsection*{Edge-to-bulk capacitive coupling $C_{\rm eb}$}

The capacitive coupling between the edge and the bulk is the most difficult contribution to evaluate. We base our estimation on  ref.~\cite{Evans1993}, which proposes a model to describe the transport in a quantum dot in the QH regime composed of a conducting island enclosed and coupled to a conducting ring. Equation (19) in ref.~\cite{Evans1993} allows to evaluate $C_{\rm eb}$ based on the charge distribution induced by a potential difference between the interfering edge channel and the bulk compressible island separated by a distance $a$. For simplicity, we assume this distance to be of the order of $l_B$ in graphene by analogy with GaAs heterostructures (see eq. (38) of ref.~\cite{Chklovskii1992} giving the width of  innermost incompressible stripe). The calculation of the capacitance also requires to set a characteristic length scale $d$ over which the influence of the potential difference is screened by nearby gate electrodes. For our device with a graphite back-gate electrode, this length scale is imposed by the thickness of the bottom hBN such that $d\simeq20~$nm. In these conditions, we can write:
 \begin{equation}
C_{\rm eb}=\frac{2L\epsilon_{\rm BN}\epsilon_0}{2\pi^2}\text{ln}\left(\frac{4d}{a}\right),
\end{equation}
which leads to $C_{\rm eb}=2.8\times10^{-17}$~F.

We expect that $C_{\rm eb}$ remains of the same order of magnitude for devices with silicon back gate because the various top gates around the FP cavity are also 20 nm away from the graphene flake and, hence, set the cutoff length $d$. More specifically, for our devices BNGr64 and BNGr30, which respectively have perimeters $2L=15.1$ and $13.3~\mu$m, we obtain $C_{\rm eb}=5.0\times10^{-17}$ and $4.4\times10^{-17}$~F. Alternatively, if we take $d=285~$nm, we get $C_{\rm eb}=10.4\times10^{-17}$ and $9.1\times10^{-17}$~F.

\subsection*{Discussion}

From these calculations, we can estimate the parameter $\xi=\frac{C_{\rm eb}}{C_{\rm b}+C_{\rm eb}}$. We obtain $\xi=0.006$ for our 3.1 $\mu$m$^2$ device with graphite back gate, confirming that oscillations should arise from pure Aharonov-Bohm effect~\cite{Halperin2011,Sivan2016}. Similarly, for our devices with silicon back gate, we obtain $\xi=0.03-0.07\ll 1$ also consistent with the observation of oscillations in the AB regime.  This analysis is fully consistent with the absence of charging effect in our graphene devices.

 \newpage

\section{Aharonov-Bohm oscillations versus dc voltage bias: asymmetry and decay}
\label{bias_osc}
 In this section, we discuss the oscillations induced by the application of a dc voltage bias and explain the origin of the tilted checkerboard pattern. We also analyze the decay of the oscillations amplitude with the voltage bias related to an energy relaxation or dephasing process.

\subsection*{Theoretical model for asymmetric potential drop}
Here, we derive the formula for the transmission of a QH-FP interferometer as a function of magnetic field and voltage bias using the same formalism as in ref.~\cite{Chamon1997}, but we take into account a possible asymmetric potential drop at the two QPCs. 

The transmission of a non-interacting QH Fabry-P\'erot interferometer reads:
\begin{equation}
 t(\epsilon,\Phi)=\frac{t_1 t_2 e^{i \pi \frac{\Phi}{\Phi_0}+i\frac{ L \epsilon}{\hbar v}}}{1-r^{'}_1 r_2 e^{2i \pi \frac{\Phi}{\Phi_0} +i\frac{ 2L\epsilon}{\hbar v} }},
\label{eq1}
\end{equation}
where $2 \pi \frac{\Phi}{\Phi_0}$ is the Aharonov-Bohm phase, $ \frac{2L\epsilon}{\hbar v}$ the dynamical phase accumulated by electrons after one winding in the cavity of length $2L$, $t_1$ and $t_2$  the transmission amplitudes of QPC$_1$ and QPC$_2$ for right moving particles, $r'_1$ the reflection amplitude for left-movers at QPC$_1$ and $r_2$ the reflection amplitude for right-movers at QPC$_2$.

The transmission probability is:
\begin{equation} 
T(\epsilon,\Phi)=\frac{\mid t_1 \mid^2 \mid t_2 \mid^2}{1+\mid r^{'}_1 r_2 \mid^2 -2 \mid r_1^{'} r_2 \mid \text{cos}(2 \pi \frac{\Phi}{\Phi_0} +\frac{ 2L\epsilon}{\hbar v}  +\varphi )},
\label{eq2}
\end{equation}
where $\varphi$ is a constant phase factor which depends on the scattering phase of the QPCs. Given that ${\mid r_{1,2} \mid}^2= {\mid r'_{1,2} \mid}^2=R_{1,2}$ and ${\mid t_{1,2} \mid}^2=T_{1,2}$, we can rewrite (\ref{eq2}) as
\begin{equation}
T(\epsilon,\Phi)=\frac{T_1 T_2}{1+ R_1 R_2 - 2\sqrt{R_1 R_2} \text{ cos}(2 \pi \frac{\Phi}{\Phi_0} +\frac{ 2L\epsilon}{\hbar v} +\varphi )}.
\label{eq3}
\end{equation}

In the weak backscattering limit, $R_i \ll 1$, and omitting the constant phase term $\varphi$, we obtain at first order:
\begin{equation}
T(\epsilon,\Phi)=1-R_1-R_2+2\sqrt{R_1 R_2}\text{ cos}\left(2 \pi \frac{\Phi}{\Phi_0} +\frac{ 2L\epsilon}{\hbar v}\right)
 \end{equation}

We then consider a finite dc voltage bias $V$ applied between source and drain contacts. We note $q=-e<0$ the electron charge. Depending on the energy relaxation processes consecutive to the current flow, and on the electrostatic coupling between the cavity, the back gate, the source and  the drain, the electrochemical potential in the cavity will adjust itself at a value intermediate between that of the source and that of the drain. The right-movers coming from the source contact have an energy $qV^+=qV(\frac{1}{2}+x)=qV\beta$ with respect to the chemical potential within FP cavity and the left-movers coming from the drain have an energy $qV^-=-qV(\frac{1}{2}-x)=-qV\overline{\beta}$. In these expressions, $x\in[-\frac{1}{2},\frac{1}{2}]$ is the voltage bias asymmetry factor.
$x=0$ corresponds to a symmetric biasing with $V^+=\frac{V}{2}$ and $V^-=-\frac{V}{2}$, meaning that the potential drop is the same across both QPCs. When $x=\frac{1}{2}$ (or equivalently $x=-\frac{1}{2}$) the bias is completely asymmetric, $V^+=V$ and $V^-=0$ (or equivalently $V^+=0$ and $V^-=-V$), the potential drop only occurs at one QPC while the FP cavity is at the same potential as one of the two contacts.

 At zero temperature, the current through the device is given by $I=\frac{q}{h}\int^{qV^+}_{qV^-}T(\epsilon,\Phi)\rm d\epsilon$. In the weak backscattering limit, it writes:
\begin{equation}
I = \frac{q}{h}\int^{qV^+}_{qV^-} \left[1-R_1- R_2 +2\sqrt{R_1 R_2}\text{ cos}\left(2 \pi \frac{\Phi}{\Phi_0} +\frac{ 2L\epsilon}{\hbar v}\right)\rm \right] d\epsilon = \it I_{\text{0}}+ I_{\text{osc}},
 \end{equation}
where $I_{\text{0}}=\frac{e^2}{h}(1-R_1-R_2)V$ is the constant part of the current and $I_{\text{osc}}$ is the oscillating part of the current which writes:
 \begin{equation}
I_{\text{osc}}=\frac{e^2}{h}2\sqrt{R_1 R_2}\frac{\hbar v}{2Lq}\left[\text{sin}\left(2\pi \frac{\Phi}{\Phi_0}+\frac{2L}{\hbar v}qV\beta\right)-\text{sin}\left(2\pi \frac{\Phi}{\Phi_0}-\frac{ 2L}{ \hbar v}qV\overline{\beta}\right)\right].
 \end{equation} \linebreak
The corresponding differential conductance is then:
  \begin{equation}
\frac{dI_{\text{osc}}}{dV}=g_{\text{osc}}\left[\beta\,\text{cos}\left(2\pi \frac{\Phi}{\Phi_0}-\frac{2L}{\hbar v}eV\beta \right)+\overline{\beta}\,\text{cos}\left(2\pi \frac{\Phi}{\Phi_0}+\frac{2L}{ \hbar v}eV\overline{\beta}\right)\right],
\label{eq7}
 \end{equation}
with $g_{\text{osc}}=\frac{e^2}{h}2\sqrt{R_1 R_2}$ and restoring $q=-e$.

When the potential drop at the constrictions is symmetrical, that is, $V^+=V/2$ and $V^-=-V/2$, we have $\beta=\overline{\beta}=\frac{1}{2}$ ($x=0$) and then:
 %
\begin{equation}
\frac{dI_{\text{osc}}}{dV}=g_{\text{osc}}\,\text{cos}\left(2\pi\frac{\Phi}{\Phi_0}\right)\text{cos}\left(2\pi \frac{L}{hv}eV\right),
\label{Conductance_symmetric}
 \end{equation}
leading to a checkerboard pattern  with a period versus bias voltage which is equal to the ballistic Thouless energy : $e\Delta V = hv/L = E_{\rm Th}$. 

If the bias is completely asymmetrical, for example when $V^+=V$ and $V^-=0$ with  $\beta=1$ and $\overline{\beta}=0$ ($x=\frac{1}{2}$), we obtain:  
 \begin{equation}
\frac{dI_{\text{osc}}}{dV}=g_{\text{osc}}\,\text{cos}\left(2\pi \frac{\Phi}{\Phi_0}-2\pi\frac{2L}{ hv}eV\right)
\label{Conductance_assymetric}
 \end{equation}
that draws a diagonal strip pattern with a period versus bias voltage (at fixed magnetic field) which is equal to half the Thouless energy. Any intermediate value of $x$ leads to a mixed pattern, that is, a tilted checkerboard as observed in our experiment. Note that the measured diagonal resistance $\delta R_\text{D}=-\frac{\rm d \it I_{\text{osc}}}{\rm d \it V }(\frac{h}{e^2 })^2$ shows exactly the same oscillatory features as the conductance in the weak backscattering limit.

In Fig.~\ref{fig:Checkerboard_nu1}, we gather the results obtained in the three different interferometers as a function of voltage bias (Fig.~\ref{fig:Checkerboard_nu1}a, c, d and f are respectively identical to Fig.~3c, d, e and f). The checkerboard patterns are tilted for our small (a) and medium interferometers (b), whereas the tilt is hardly visible for the largest interferometer (c). Using eq. (\ref{eq7}), we can quantitatively reproduce in Fig.~\ref{fig:Checkerboard_nu1}d, e and f the three experimental checkerboards with asymmetry parameters  $x=0.2$, 0.1 and 0.02, respectively.

In our experiment, we apply a dc voltage to the source contact while the drain contact is kept grounded. The electrostatic coupling of the cavity to the back-gate electrode results in an asymmetric potential drop which could explain why the checkerboard patterns of our two smallest interferometers are tilted. On the other hand, the fact that the checkerboard pattern is nearly symmetric for the largest interferometer, indicates that energy relaxation processes equilibrate the chemical potential for sufficiently large interferometers, leading to a symmetric potential drop. Interestingly, tilted checkerboards in QH-FP interferometers has never been reported for GaAs QH-FP devices of the same size as our small interferometer, possibly due to the larger back-gate coupling in our graphene device equipped with a graphite back gate, or because the chemical potential equilibration is less effective in graphene.

\subsection*{Decay of the oscillations at finite bias}

For an asymmetric potential drop characterized by an asymmetry factor $x$, the amplitude of the flux-periodic oscillations given by eq.~(\ref{eq7}) oscillates versus bias voltage with the following dependence:
\begin{equation}
\mathcal{A}\left(V,E_{\rm Th}/e\right) = \sqrt{ \cos^2\left(2\pi\frac{eV}{E_{\rm Th}}\right) + 4x^2 \sin^2\left(2\pi\frac{eV}{E_{\rm Th}}\right) }
\label{equation:amplitude-vs-bias}
\end{equation}

Note that the period of this function is always the Thouless energy $E_{\rm Th}=hv/L$ whatever the asymmetry factor $x$, whereas the period of the conductance oscillations versus bias voltage at fixed magnetic field varies with the value of $x$ (see for example eq.~(\ref{Conductance_symmetric}) and eq.~(\ref{Conductance_assymetric})\,).

\begin{figure*}[h!]
 \centering
 \includegraphics[width=1\textwidth]{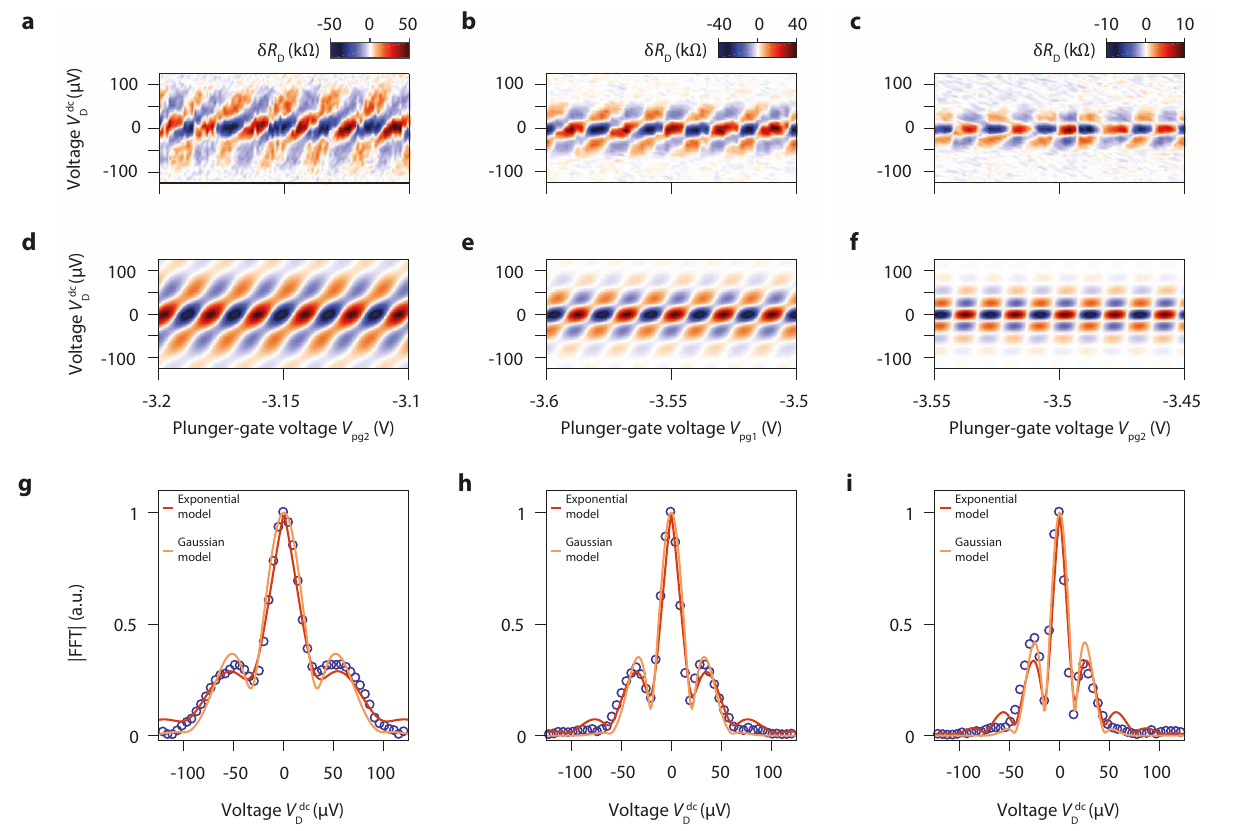}
\caption{\textbf{Bias dependence of Aharonov-Bohm oscillations.} \textbf{a, b, c,} Differential diagonal resistance variations $\delta R_D$, after background subtraction, versus dc  diagonal voltage $V^{\text{dc}}_{\text{D}}$ and plunger-gate voltage $V_{\text{pg1,2}}$ for the small, medium and large interferometer respectively in a, b and c. Interferences are obtained with the outer edge channel at 14 T. \textbf{e, f, g,} Numerical simulations of resistance oscillations induced by voltage bias and plunger-gate voltage that reproduce the data presented in a, b and c, respectively. The simulations incorporate an asymmetric potential drop at the two QPCs and an out-of-equilibrium decoherence factor. The voltage bias asymmetry factors of $x=0.2$  and $x=0.1$,  respectively for the small d and medium interferometer e, are significant, indicating a limited chemical potential equilibration as opposed to the large interferometer f, which has a very small asymmetry term $x=0.02$. \textbf{g, h, i,} Amplitude of the Fourier transform of the oscillations at fixed voltage bias (blue dots) and fits with eq. (\ref{eq10_bis}) (red line) and  eq. (\ref{eq_gauss_decay_bis}) (orange line). Fitting parameters are reported in Table \ref{Table_fit_bias}.}
\label{fig:Checkerboard_nu1}
\end{figure*}

In Fig.~\ref{fig:Checkerboard_nu1}a, b and c, however, we observe that the oscillations amplitude decays rapidly with the bias voltage and vanishes typically after one voltage period. Such a fast decay is much faster than the $1/\Delta V$ dependence predicted in ref.~\cite{Chamon1997} and was already reported by McClure and coworkers~\cite{McClure2009} in GaAs QH-FP interferometers. These authors found that an exponential decay of the oscillations amplitude with the bias describes correctly the data. Theoretical investigations~\cite{NgoDinh2012} confirmed that Coulomb interactions can lead to an approximate exponential decay. Following this approach, we fitted the oscillations in our data with:
\begin{equation}
\mathcal{A}(V,\Delta V_{\rm expo}) \, \exp\left(-2\pi\chi\frac{\vert V \vert}{{\Delta V_{\rm expo}}}\right),
\label{eq10_bis}
\end{equation}
where $\chi$ is a phenomenological parameter that describes how fast the oscillations vanish with voltage, and $\Delta V_{\rm expo}$ is the period of the resistance oscillations for this exponential decay. The amplitude of the oscillations is obtained by computing the Fourier amplitude of the resistance oscillations as a function of the plunger-gate voltage at fixed bias voltage. This leads to the lobe structure shown in Fig.~\ref{fig:Checkerboard_nu1}g, h and i. A good agreement between the model and the data is found for the three interferometers. The extracted voltage periods $\Delta V_{\rm expo}$ and damping factors $\chi$ are reported in Table~\ref{Table_fit_bias}. It is worth noticing, however, that this phenomenological model does not capture the absence of secondary lobes in the experiments, suggesting that the decay of the oscillations is faster than exponential.

We therefore consider a second model with a Gaussian decay of the bias-induced oscillations. Investigations in Mach-Zehnder interferometers revealed that a Gaussian decay may arise from phase fluctuations of the interfering edge channel due to Coulomb interactions or the electric noise in the non-interfering edge channels~\cite{Roulleau2007,Roulleau2008b,Litvin08,Yamauchi2009}. Within this approach, we fitted our data with:
\begin{equation}
\mathcal{A}(V,\Delta V_{\rm gauss}) \, \exp\left(-\frac{ V^2}{{2V_{0}}^2}\right),
\label{eq_gauss_decay_bis}
\end{equation}
where $V_{0}$ is the voltage scale characterizing the width of the Gaussian envelope, and $\Delta V_{\rm gauss}$ the period of the resistance oscillation for this Gaussian decay. The fits of the experimental data with this expression are displayed in Fig.~\ref{fig:Checkerboard_nu1}g, h and i (orange lines). This second model also describes well the data. The extracted voltage periods $\Delta V_{\rm gauss}$, reported in Table~\ref{Table_fit_bias}, are close to those obtained with the exponential decay model. The extracted $V_{0}$ values scale linearly with the inverse interfering path length $1/L$ as mentioned in ref.~\cite{Yamauchi2009} and is typically one third of $\Delta V_{\rm gauss}$.

The qualitative difference between the exponential and Gaussian decays is that the exponential decay fits better the amplitude of the first lobe but fails to reproduce the vanishing of the second ones, whereas the Gaussian model is less accurate for the first lobe but shows a suppressed second lobe. 

\begin{table*}[h!]
\centering
\begin{tabular}{|l|c|c|c|c|}
\hline
  QH-FP  & $\Delta V_{\rm expo}$ ($\mu \text{V}$) & $\chi$ & $\Delta V_{\rm gauss}$ ($\mu \text{V}$) & $V_0$ ($\mu \text{V}$)\\
  
  \hline \hline
  Small & 134  & 0.42 & 128 & 40\\
  \hline  Medium & 83 & 0.42 & 81 & 25\\
  \hline   Large &57& 0.35 & 61 & 21\\
  \hline   
\end{tabular}

    \caption{\textbf{Fitting parameters for the different models of bias-induced oscillation decay.} Voltage period $\Delta V_{\rm expo}$ for the exponential decay model; $\chi$ damping rate for the exponential decay model; voltage period $\Delta V_{\rm gauss}$ for the Gaussian decay model; $V_0$ width of the Gaussian envelope.}
    \label{Table_fit_bias}
\end{table*}

\newpage
\clearpage

\newpage
\clearpage

\section{Temperature dependence of the Aharonov-Bohm oscillations: thermal averaging}

The effect of temperature on the visibility of the Aharonov-Bohm oscillations has been calculated in ref.~\cite{Chamon1997} in the limit of weak backscattering and at finite bias voltage for a symmetric potential drop at the two constrictions. This calculation considers only the thermal averaging of the interference and does not introduce decoherence by inelastic scattering or energy relaxation at finite bias.

Here we explain in details the calculation in the symmetric case and then extend the result to the case of an asymmetric potential drop as observed in our device. In the following, we use the expression of the transmission coefficient obtained in the previous section in the limit of weak backscattering:
\begin{equation}
T(E,\Phi) = 1 - R_1 - R_2 + \sqrt{R_1R_2} \left( e^{i2\pi\Phi/\Phi_0} e^{i E 2L / \hbar v} + e^{-i2\pi\Phi/\Phi_0} e^{-i E 2L / \hbar v} \right)
\end{equation}

\subsection*{Symmetric potential drop}
Assuming a symmetric potential drop at the two constrictions as in ref.~\cite{Chamon1997}, the current at finite temperature $T$ and finite voltage $V$ is given by:
\begin{equation}
I(\Phi,V,T) = \frac{q}{h} \int_{-\infty}^{+\infty} T(E,\Phi) \left( \frac{1}{1+e^{(E-\frac{qV}{2})/k_{\rm B}T}} - \frac{1}{1+e^{(E+\frac{qV}{2})/k_{\rm B}T}} \right) {\rm d}E,
\end{equation}
where $q<0$ is the electron charge. Using the expression of the transmission coefficient in the limit of weak back-scattering, the current writes:
\begin{equation}
I(\Phi,V,T) = \frac{q^2}{h} (1-R_1-R_2) V - \frac{q}{h} \sqrt{R_1R_2} \left(e^{i2\pi\Phi/\Phi_0}H(V,T)+e^{-i2\pi\Phi/\Phi_0}H(V,T)^*\right),
\end{equation}
where we introduce the function:
\begin{equation}
H(V,T) = \int_{-\infty}^{+\infty} e^{i E 2L / \hbar v} \left( \frac{1}{1+e^{(E-\frac{qV}{2})/k_{\rm B}T}} - \frac{1}{1+e^{(E+\frac{qV}{2})/k_{\rm B}T}} \right) {\rm d}E.
\end{equation}
By changing the variable in the integral, it becomes:
\begin{equation}
H(V,T) = \left( e^{i \frac{qV}{2} 2L / \hbar v} - e^{-i \frac{qV}{2} 2L / \hbar v} \right) \int_{-\infty}^{+\infty} e^{i E 2L / \hbar v} \, \frac{1}{1+e^{E/k_{\rm B}T}} \, {\rm d}E,
\end{equation}
where the choice of a symmetric potential drop influences only the term in the parenthesis. The calculation of the integral gives:
\begin{equation}
\int_{-\infty}^{+\infty} e^{i E 2L / \hbar v} \, \frac{1}{1+e^{E/k_{\rm B}T}} \, {\rm d}E = -i 2\pi k_{\rm B}T \, \sum_{n=0}^{+\infty} e^{-\omega_n 2L / \hbar v} = \frac{-i 2\pi k_{\rm B}T}{2 \sinh(\pi k_{\rm B}T 2L / \hbar v)},
\end{equation}
where $\omega_n=(2n+1)\pi k_{\rm B}T$ are the Matsubara frequencies, with $n\in\mathbb{Z}$. In this case of a symmetric potential drop, the function $H(V,T)$ is real and writes:
\begin{equation}
H(V,T) = \sin(qV L / \hbar v) \, \frac{2\pi k_{\rm B}T}{\sinh(\pi k_{\rm B}T 2L / \hbar v)}.
\end{equation}
The current finally writes:
\begin{equation}
I(\Phi,V,T) = G_0 V - \frac{q}{h} \, \sqrt{R_1R_2} \,\,\, 2 \, \cos(2\pi\Phi/\Phi_0) \, \sin(qV L / \hbar v) \, \frac{2\pi k_{\rm B}T}{\sinh(\pi k_{\rm B}T 2L / \hbar v)},
\end{equation}
which is equivalent to equations (16) and (18) in ref.~\cite{Chamon1997}. The differential conductance writes:
\begin{equation}
G(\Phi,V,T) = G_0 - \frac{q^2}{h} \, \sqrt{R_1R_2} \,\,\, 2 \, \cos(2\pi\Phi/\Phi_0) \, \cos(qV L / \hbar v) \, \frac{\pi k_{\rm B}T 2L / \hbar v}{\sinh(\pi k_{\rm B}T 2L / \hbar v)},
\label{Tdep_full}
\end{equation}
which forms a checkerboard pattern as a function of field and voltage. At high temperature, the visibility of these oscillations decreases exponentially with a dependence of the form:
\begin{equation}
e^{-\pi k_{\rm B}T 2L / \hbar v} = e^{-4\pi^2 k_{\rm B}T / E_{\rm Th}} = e^{-T/T_0},
\end{equation}
where $E_{\rm Th}=hv/L$ is the ballistic Thouless energy which corresponds to the oscillation period $q\Delta V$ versus bias voltage, and $T_0$ is the fitting parameter of the exponential temperature dependence which is related to the Thouless energy by:
\begin{equation}
4\pi^2 k_{\rm B} T_0 = E_{\rm Th} = q\Delta V.
\end{equation}

\subsection*{Asymmetric potential drop}

In case of an asymmetric potential drop at the two constrictions (see section~\ref{bias_osc}), the potential energy is $qV^+={\beta}qV$ at the source contact and $qV^-=-\bar{\beta}qV$ at the drain contact, with $\beta=\frac{1}{2}+x$ and $\bar{\beta}=\frac{1}{2}-x$ with the parameter $x\in[-\frac{1}{2},\frac{1}{2}]$ characterizing the asymmetry of the potential drop. The current at finite temperature $T$ and finite voltage $V$ is then given by:
\begin{equation}
I(\Phi,V,T) = \frac{q}{h} \int_{-\infty}^{+\infty} T(E,\Phi) \left( \frac{1}{1+e^{(E-{\beta}qV)/k_{\rm B}T}} - \frac{1}{1+e^{(E+\bar{\beta}qV)/k_{\rm B}T}} \right) {\rm d}E
\end{equation}
Following the same calculations as above now gives the function:
\begin{equation}
H(V,T) = e^{i x qV 2L / \hbar v} \sin(qV L / \hbar v) \, \frac{2\pi k_{\rm B}T}{\sinh(\pi k_{\rm B}T 2L / \hbar v)}
\end{equation}
which contains a complex phase factor. The current writes:
\begin{equation}
I(\Phi,V,T) = G_0 V - \frac{q}{h} \, \sqrt{R_1R_2} \,\,\, 2 \, \cos(2\pi\Phi/\Phi_0 + x qV 2L / \hbar v) \, \sin(qV L / \hbar v) \, \frac{2\pi k_{\rm B}T}{\sinh(\pi k_{\rm B}T 2L / \hbar v)}
\end{equation}
which is modified only by the term $x qV 2L / \hbar v$ in the cosine function. The differential conductance writes:
\begin{equation}
G(\Phi,V,T) = G_0 - \frac{q^2}{h} \, \sqrt{R_1R_2} \,\,\, 2 \, g(\Phi,V) \, \frac{\pi k_{\rm B}T 2L / \hbar v}{\sinh(\pi k_{\rm B}T 2L / \hbar v)}
\end{equation}
where the oscillation term:
\begin{equation}
g(\Phi,V) = \cos(2\pi\Phi/\Phi_0 + x qV 2L / \hbar v) \cos(qV L / \hbar v) - 2x \sin(2\pi\Phi/\Phi_0 + x qV 2L / \hbar v) \sin(qV L / \hbar v)
\end{equation}
gives a titled checkerboard pattern as a function of field and voltage for $x\neq0$. It is interesting to note that the temperature dependence is not affected by the asymmetry of the potential drop at the constrictions. The fitting parameter $T_0$ of the exponential temperature dependence is still related to the ballistic Thouless energy by $4\pi^2 k_{\rm B} T_0 = E_{\rm Th}$. 

\newpage

\section{Evaluation of the phase coherence length $L_{\phi}$}

To estimate the phase coherence length $L_{\phi}$ in our graphene QH-FP interferometers, we assume that the visibility $\mathcal{V}$ of coherent oscillations scales as:
\begin{equation} 
	\mathcal{V}=\mathcal{V}_0\frac{2L /L_T}{\sinh(2L/L_T)}\exp\left(-\frac{2L}{L_{\rm \phi}(T)}\right)
\label{Visibility_with_thermal_averaging}
\end{equation} 
where $L_{T}=\frac{hv}{2\pi^2k_{B}T}$ is the characteristic length associated with the decay of the visibility due to thermal averaging at temperature $T$ (see eq. (\ref{Tdep_full}) in previous section), $L_{\phi}(T)$ is the phase coherence length that can depend on temperature, $2L$ is the the perimeter of the FP cavity and $\mathcal{V}_0$ is the asymptotic limit reached by the visibility when L tends to zero. Note that the exponential decrease due the finite coherence length is only valid for $2L$ above $L_{\phi}$ and should saturate to a particular visibility below unity for smaller perimeters.

Fitting the evolution of $\mathcal{V}$ with $2L$ at fixed temperature with eq. (\ref{Visibility_with_thermal_averaging}) provides a direct estimate of $L_{\phi}$. As visibility depends on the QPC transmissions, we performed this length-dependence analysis by considering our best visibility data obtained for the three sizes of interferometers at 14 T. We evaluate the electron temperature at our base fridge temperature to be $T\simeq 20~ \rm mK$, which corresponds to the temperature below which the $T$-dependence of the visibility saturates. For experiments with the inner edge channel, we extracted the visibility through $\frac{G_{\rm max}-G_{\rm min}}{(G_{\rm max}-e^2/h)+(G_{\rm min}-e^2/h)}$, which subtracts the conductance contribution of the fully transmitted outer edge channel. 

\begin{figure}[h!]
 \centering
 \includegraphics[width=1\textwidth]{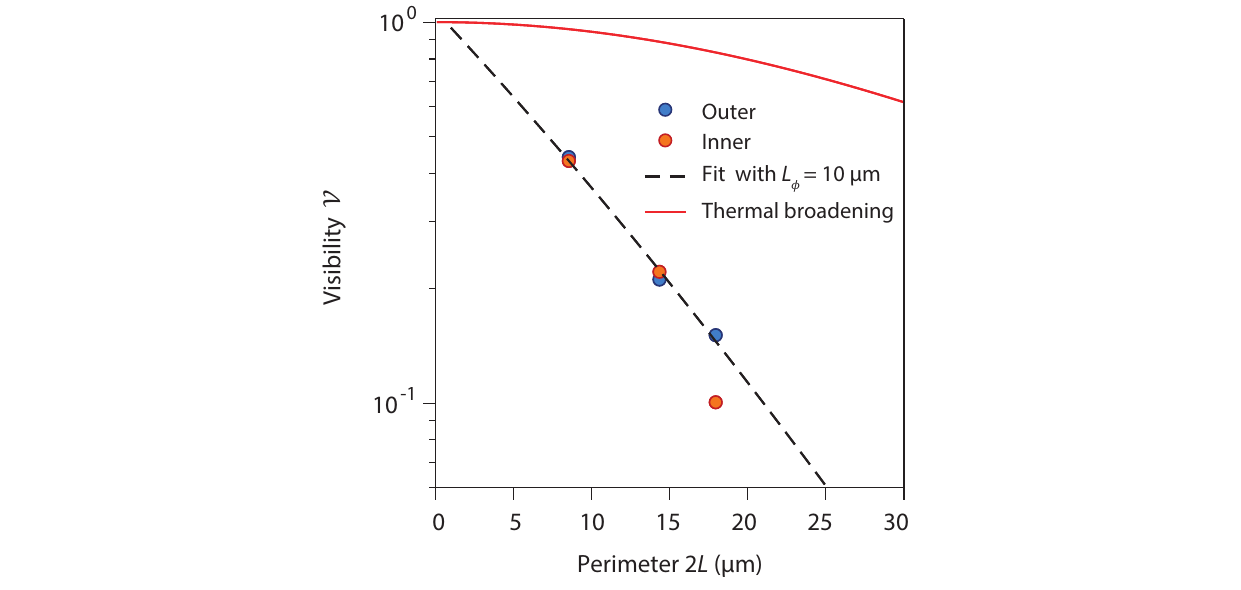}
\caption{\textbf{Phase coherence length $L_{\phi}$.} Evolution of the best visibilities $\mathcal{V}$  with the perimeter $2L$ of the interferometers obtained in experiments at base temperature with the outer (blue dots) and the inner (red dots) edge channel. The red solid line shows the thermal broadening contribution.  The fit of the data (black dashed line) with eq. (\ref{Visibility_with_thermal_averaging}) and discarding the inner edge channel experiment for the large interferometer, provides a coherence length of 10 $\mu$m at 20 mK.}
\label{fig:L_phi}
\end{figure}

Fig.~\ref{fig:L_phi} shows the evolution of these visibilities $\mathcal{V}$ with the perimeter of the interferometers 2$L$. For comparison, the decrease of the visibility induced by the thermal broadening at 20 mK is also shown with the solid red line (eq. (\ref{Visibility_with_thermal_averaging}) with $L_{\rm \phi}(T)$ infinite and a edge state velocity of $1.4\times10^{5}$~m/s, giving $L_T=17~\mu$m). For both experiments with the outer and the inner edge channel, a fast decrease of $\mathcal{V}$ with $2L$ is observed which cannot be explained by the effect of thermal broadening. The best visibilities for both interfering edge channels are virtually the same except for data in the large interferometer with the inner edge channel, which shows a significant drop compared to the data with the outer one. It probably reflects that the tuning of the QPC could have been improved. We thus discard it for our quantitative analysis.

By fitting the visibility decay, we extract a phase coherence length $L_{\phi}\approx 10~\mu$m at 20 mK and 14 T. The obtained value of 10 $\mu$m is smaller or comparable to the perimeter length, which justifies the exponential decrease used in eq.~(\ref{Visibility_with_thermal_averaging}) (the saturation would appear for smaller perimeters as the ones studied here). This value is also consistent with the observation of coherent Aharonov-Bohm oscillations in the double FP cavity at base temperature.

\newpage
\section{Analysis of the double-cavity interferometer}

 In this section we discuss the experiments performed in the coherently-coupled double FP cavity. We first derive the theoretical expression for the transmission for a double cavity and then compare it with our data to show that electron transport remains coherent in the overall device.


 The transmission and reflection amplitudes of a Fabry-P\'{e}rot interferometer reads:
\begin{equation}
 t_{\text{FP}}(\varphi)=\frac{t_1 t_2 e^{i\varphi}}{1-r^{'}_1 r_2 e^{i 2\varphi}},
\end{equation}
\begin{equation}
t'_{\text{FP}}(\varphi)=\frac{t'_1 t'_2 e^{i\varphi}}{1-r^{'}_1 r_2 e^{i 2\varphi}},
\end{equation}
\begin{equation}
r_{\text{FP}}(\varphi)=r_1+\frac{r_2 t_1 t'_1 e^{i 2\varphi}}{1-r^{'}_1 r_2 e^{i 2\varphi}},
\end{equation}
\begin{equation}
r'_{\text{FP}}(\varphi)=r'_2 +\frac{r'_1 t'_2 t_2 e^{i 2\varphi}}{1-r^{'}_1 r_2 e^{i 2\varphi}},
\end{equation}
where $2\varphi$ is the Aharonov-Bohm phase accumulated by electrons after one winding in the cavity, $t_i$ ($t'_i$) the transmission amplitude, and $r_i$ ($r'_i$) the reflection amplitude of QPC$_i$ for right (left) moving particles.

The total transmission amplitude $t_{\text{tot}}$ of two coupled FP cavities can be calculated using the transmission and reflection amplitudes of one FP cavity and the transmission and reflection amplitudes of a third QPC. Thus, using the previous expressions, we have:
\begin{equation}
 t_{\text{tot}}(\varphi_1,\varphi_2)=\frac{t_{\text{FP}}(\varphi_1) t_3 e^{i\varphi_2}}{1-r^{'}_{\text{FP}}(\varphi_1) r_3 e^{i2\varphi_2}},
\end{equation}
where $2\varphi_{1}$ and $2\varphi_{2}$ are the Aharonov-Bohm phase accumulated by electrons after one winding in the cavity between QPC$_1$ and QPC$_2$ and between QPC$_2$ and QPC$_3$, respectively.

Using $\vert t_i \vert^2=\vert t'_i \vert^2= T_{i}$, $\vert r_i \vert^2=\vert r'_i \vert^2= R_{i}$ and the relation $r_i'=-\bar{r_i}t'_i/\bar{t_i}$ (the overline indicates complex conjugate), we can express the transmission as:
\begin{equation} 
T_{\text{tot}}(\phi_1,\phi_2)=\frac{T_1 T_2 T_3}{\vert 1-\sqrt{R_1 R_2}e^{i\phi_1}-\sqrt{R_2 R_3}e^{i\phi_2}+ \sqrt{R_1 R_3}e^{i(\phi_1+\phi_2)}\vert^2}=\frac{T_1 T_2 T_3}{D},
\end{equation}
where $\phi_{1}$ and $\phi_{2}$ are the Aharonov-Bohm phases acquired when quasiparticles wind into the medium and small cavities respectively (including the phase factor from the reflection amplitudes of the QPCs). The denominator $D$ can be written as:
\begin{equation}
\begin{multlined}
  D = 1+ R_1 R_2 + R_3 R_2 +R_1 R_3
 -2(1+R_3)\sqrt{R_1 R_2}\cos(\phi_1)-2(1+R_1)\sqrt{R_2 R_3}\cos(\phi_2)
\\+2 \sqrt{R_1 R_3}\cos(\phi_1+\phi_2)+2R_2 \sqrt{R_1 R_3}\cos(\phi_1-\phi_2).
\end{multlined}
\label{Denominatorsum}
\end{equation}

In this expression, four oscillation frequencies emerge, namely, $\phi_1$, $\phi_2$, $\phi_3=\phi_1 + \phi_2$ and  $\phi_4=\phi_1 - \phi_2$. The terms in $\phi_3$ and $\phi_4$ in eq. (\ref{Denominatorsum}), which result from coherent interferences through the two interferometers, does not have the same prefactor : the amplitude of the $\phi_3$ oscillations is larger than the amplitude of the $\phi_4$ oscillations which is even negligible in the weak backscattering limit. In contrast, in a situation where the transport through the double cavity would be incoherent, one could expect the appearance of term in the form of $\cos(\phi_1)\times \cos(\phi_2)=\frac{1}{2} \left[ \cos(\phi_3)+ \cos(\phi_4) \right]$ which would lead to equal amplitudes of  $\phi_3$ and $\phi_4$ oscillating components.

%

Relating this model to our device geometry, we can ascribe to each of these four Aharonov-Bohm phases a coupling to the relevant plunger gates: 
  \begin{equation}
 \phi_1\simeq\frac{2\pi}{\Phi_0}(\delta A_{1}B+A_{1}\delta B)=\frac{2\pi}{\Phi_0}(\alpha_1 V_{\text{pg1}}B+A_{1}\delta B),
\end{equation}
   \begin{equation}
\phi_2\simeq\frac{2\pi}{\Phi_0}(\delta A_{2}B+A_{2}\delta B)=\frac{2\pi}{\Phi_0}(\alpha_2 V_{\text{pg2}}B+A_{2}\delta B),
   \end{equation}
   \begin{equation}
\phi_3\simeq\frac{2\pi}{\Phi_0}\left[(\delta A_{1}+\delta A_{2})B+(A_{1}+A_{2})\delta B\right]=\frac{2\pi}{\Phi_0}\left[(\alpha_1 V_{\text{pg1}}+\alpha_2 V_{\text{pg2}})B+(A_{1}+A_{2})\delta B\right],
   \end{equation} 
   \begin{equation}
\phi_4\simeq\frac{2\pi}{\Phi_0}\left[(\delta A_{1}-\delta A_{2})B+(A_{1}-A_{2})\delta B\right]=\frac{2\pi}{\Phi_0}\left[(\alpha_1 V_{\text{pg1}}-\alpha_2 V_{\text{pg2}})B+(A_{1}-A_{2})\delta B\right],
   \end{equation}  
where $A_{1}$ and $A_{2}$ are the area of the medium and small cavities, respectively, $V_{\text{pg1}}$ and $V_{\text{pg2}}$ the plunger-gate voltages that tune these areas and $\alpha_{1}$ and $\alpha_{2}$ their lever arms.

   \begin{figure}[h!]
  	\centering
  	\includegraphics[width=1\textwidth]{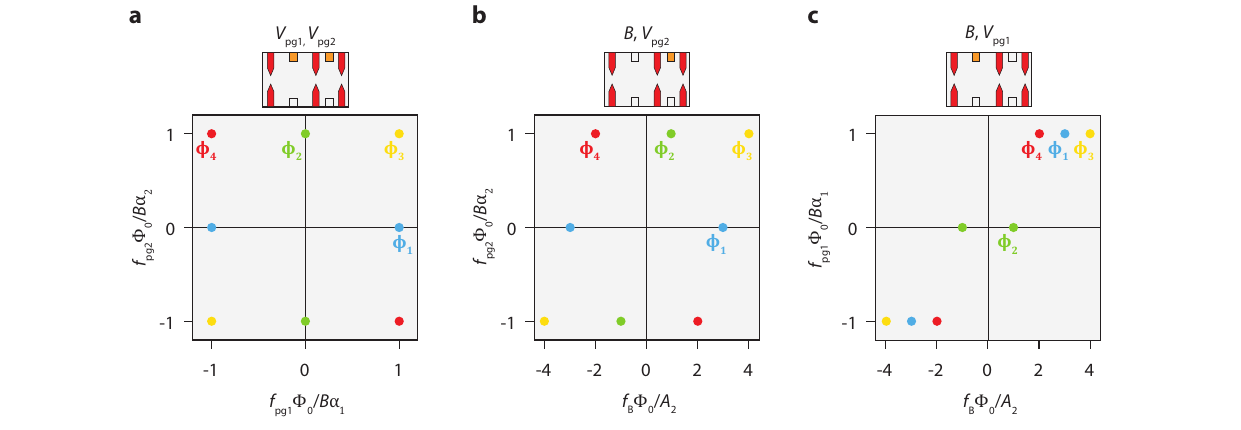}
  	\caption{\textbf{Fourier analysis of double QH-FP interferometer.} \textbf{a, b, c,} Positions in reciprocal space of the oscillation frequencies for three different configurations of  interferometry experiments (assuming $A_1=3A_2$). Each peak is labelled with its Aharonov-Bohm phase. Top schematics depict the active QPCs (red) and plunger gates (orange) in each experiments. The parameters used to tune the Aharonov-Bohm phases in each case are indicated above the corresponding schematic.}
  	  	\label{fig:simu_norm}
  \end{figure}
 
In Figure~\ref{fig:simu_norm} we show the expected frequencies in Fourier space for a coherently-coupled double QH-FP interferometer upon varying both plunger gates (Fig.~\ref{fig:simu_norm}a), or one plunger gate and the magnetic field (Fig.~\ref{fig:simu_norm}b and c). For the former case, the plunger-gate frequencies corresponding to the small and medium interferometers are located on the $x$ and $y$ axis, reflecting the terms $\phi_1$ and $\phi_2$ in eq. (\ref{Denominatorsum}), whereas the double interferometer terms $\phi_3$ and $\phi_4$ that depend on both plunger gates are located on the diagonals. For latter configurations, the frequency of the interferometer without the active plunger gate depends only on $B$ and is thus located at zero plunger-gate frequency on the horizontal axis ($\phi_1$ in b and $\phi_2$ in c), whereas the frequency of the other interferometer with the active plunger gate, as well as the coupled interferometer frequencies, are located at finite plunger gate frequency. 

\begin{figure*}[h!]
  	\centering
  	\includegraphics[width=\textwidth]{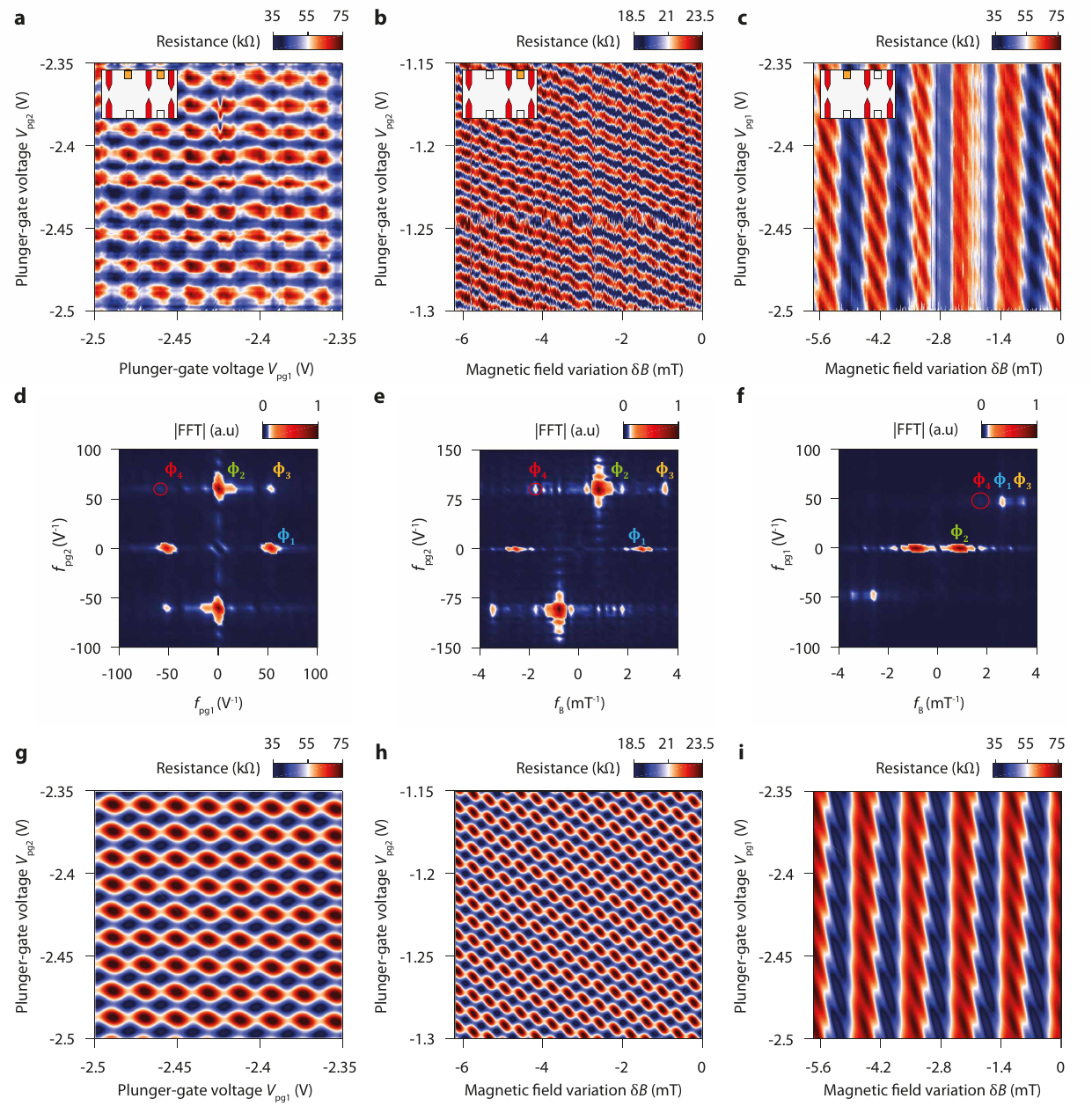}
  	\caption{\textbf{Coherently-coupled double QH-FP interferometer.} \textbf{a,} Diagonal resistance versus plunger-gate voltages $V_{\rm pg1}$ and $V_{\rm pg2}$ (outer edge channel interfering, $B$ = 14 T). \textbf{b,} Diagonal resistance versus magnetic field variation $\delta B$ and plunger-gate voltage $V_{\rm pg2}$ (inner edge channel interfering, 
$B$ = 14 T). \textbf{c,} Diagonal resistance versus magnetic field variation $\delta B$ and plunger-gate voltage $V_{\rm pg1}$ (outer edge channel interfering, $B$ = 14 T). The inset schematics in a,  b and c indicate the active QPCs (in red) and plunger gates (in orange) for the respective measurements. a and b  are identical to the Fig. 4c and 4d of the main text. \textbf{d, e, f,} Four-quadrant Fourier amplitude of the resistance oscillations displayed respectively in a, b and c in their respective reciprocal space. The peaks corresponding to the different Aharonov-Bohm phases are identified in each case. \textbf{g, h, i,} Numerical simulations reproducing the experiments shown respectively in a, b and c with eq. (\ref{Simu_3FPI}). The parameters ($\delta R_{1}$, $\delta R_{2}$, $\delta R_{3}$) are (0.66, 1, 0.18) in g, (0.64, 1, 0.22) in h, and (0.19, 1, 0.11) in i.}
  	  	\label{fig:DoubleInterfero}
  \end{figure*}

In Figure~\ref{fig:DoubleInterfero} we reproduce the data shown in Fig. 4 for the coherently-coupled QH-FP interferometer and add the configuration with $V_{\rm pg1}$ active and magnetic field (Fig.~\ref{fig:DoubleInterfero}c), which provides another confirmation of the presence of the $\phi_3$ contribution.  The four quadrants of the Fourier amplitudes are shown in order to check the presence of the $\phi_4=\phi_1 - \phi_2$ frequency. The $\phi_4$ frequency, whose expected location is indicated by the red circle in Fig.~\ref{fig:DoubleInterfero}d-f, is clearly present in the configuration of Fig.~\ref{fig:DoubleInterfero}e. Its amplitude is smaller than the amplitude of the $\phi_3$ contribution as expected in eq. (\ref{Denominatorsum}). For the two other configurations, this $\phi_4$ frequency is hardly visible. This detailed analysis provides compelling evidence for coherent transport through the three QPCs.

We can furthermore simulate the data by a simplified model that neglects terms in $R^2$ in eq.~(\ref{Denominatorsum}):
\begin{equation} 
\delta R=\delta R_{1}\cos(\phi_1)
+\delta R_{2}\cos(\phi_2)+\delta R_{3}\cos(\phi_3).
\label{Simu_3FPI}
\end{equation}
Using the experimental Fourier amplitudes for the parameters $\delta R_{1}$, $\delta R_{2}$ and $\delta R_{3}$ we obtain the resistance maps shown in Fig.~\ref{fig:DoubleInterfero}g-i that reproduce the experimental maps in Fig.~\ref{fig:DoubleInterfero}a-c with excellent fidelity. 
\newpage

\newpage
\clearpage

\end{document}